\documentclass[reprint,amsfonts, amssymb, amsmath,  showkeys,pra,nofootinbib, twocolumn, superscriptaddress,longbibliography,aps]{revtex4-2}
\usepackage{graphicx}
\usepackage{amsfonts}
\usepackage[dvipsnames]{xcolor}
\usepackage{comment}
\usepackage{xcolor}
\usepackage[caption=false]{subfig}
\usepackage{amssymb}
\usepackage{acronym}
\usepackage{soul}
\usepackage[hidelinks]{hyperref} 
\usepackage{braket}
\usepackage{tabularx}
\usepackage{multirow}
\usepackage{cleveref}   
\usepackage{svg}
\usepackage[normalem]{ulem}
\usepackage{bbold}

\definecolor{granata}{RGB}{128, 0, 32}

\makeatletter
\newcommand{\fmarki}{a}
\newcommand{\fmarkii}{b}
\newcommand{\fmarkiii}{c}
\newcommand{\fmarkiv}{d}
\newcommand{\fmarkv}{e}
\newcommand{\fmarkvi}{f}
\newcommand{\fmarkvii}{g}
\newcommand{\fmarkviii}{h}
\newcommand{\fmarkix}{i}
                
\def\@fnsymbol#1{{\ifcase#1\or \fmarki\or \fmarkii\or \fmarkiii\or \fmarkiv\or \fmarkv\or \fmarkvi\or \fmarkvii\or \fmarkviii\or \fmarkix \else\@ctrerr\fi}}
\let\switch@array\relax
\makeatother

\begin{document}

\title{Quantum Diffusion Models for Medical Image Analysis}

\author{Francesco Aldo Venturelli}
\email{francescoaldo.venturelli@upf.edu}
\affiliation{BCN Medtech, Department of Engineering, Universitat Pompeu Fabra, Barcelona, Spain,}
\affiliation{Barcelona Supercomputing Center, Barcelona, Spain,}

\author{Stefano Martina}
\email{stefano.martina@unifi.it}
\thanks{Corresponding author}
\affiliation{Department of Physics and Astronomy, University of Florence, Sesto Fiorentino, Florence, Italy,}
\affiliation{LENS - European Laboratory for Non-Linear Spectroscopy, University
of Florence, Sesto Fiorentino, Florence, Italy,}

\author{Marco Parigi}
\email{marco.parigi@unifi.it}
\affiliation{Department of Physics and Astronomy, University of Florence, Sesto Fiorentino, Florence, Italy,}
\author{Filippo Caruso}
\email{filippo.caruso@unifi.it}
\affiliation{Department of Physics and Astronomy, University of Florence, Sesto Fiorentino, Florence, Italy,}
\affiliation{LENS - European Laboratory for Non-Linear Spectroscopy, University
of Florence, Sesto Fiorentino, Florence, Italy,}

\author{Alba Cervera-Lierta}
\email{alba.cervera@bsc.es}
\affiliation{Barcelona Supercomputing Center, Barcelona, Spain,}

\author{Miguel A. González Ballester.}
\email{ma.gonzalez@upf.edu}
\affiliation{BCN Medtech, Department of Engineering, Universitat Pompeu Fabra, Barcelona, Spain,}
\affiliation{Barcelona Supercomputing Center, Barcelona, Spain,}
\affiliation{ICREA, Barcelona, Spain.}

\begin{abstract}
    Quantum Machine Learning is a novel field of research aimed at devising machine learning approaches exploiting principles of quantum mechanics, such as superposition, entanglement and interference.
    In this context, we present a scalable hybrid Quantum Diffusion Model, and evaluate its use for medical image analysis.
    Specifically, our method is based on a Discrete-Time Quantum Walk algorithm, executed on a real quantum device, to model the forward dynamics of the diffusion model. For the backward step of the diffusion model, we devise and evaluate a classical learning model, which is used to reversely denoise the data.
    In contrast with other existing attempts at applying quantum machine learning for image analysis tasks, severely limited by the size of existing quantum devices, our method allows to process real-world large size medical data. In particular, we present results on grayscale and RGB images, as well as 3D volumes of moderate sizes.
    We benchmark our results by reproducing an alternative classical counterpart model, based on diffusion models on discrete state spaces. By doing so, we compare the generation capabilities of both models in terms of three distinct state-of-the-art metrics in the field of image generation, showing the competitive, promising results of our approach.
    
\end{abstract}

\maketitle

\acrodef{ai}[AI]{Artificial Intelligence}
\acrodef{ml}[ML]{Machine Learning}
\acrodef{dl}[DL]{Deep Learning}
\acrodef{dm}[DM]{Diffusion Model}
\acrodef{genai}[GenAI]{Generative AI}
\acrodef{qgenai}[QGenAI]{Quantum Generative AI}
\acrodef{qgan}[QGAN]{Quantum Generative Adversarial Network}
\acrodef{qvae}[QVAE]{Quantum Variational Autoencoders}
\acrodef{gan}[GAN]{Generative Adversarial Networks}
\acrodef{ddpm}[DDPM]{Denoising Diffusion Probabilistic Model}
\acrodef{kl}[KL]{Kullback-Leibler}
\acrodef{fid}[FID]{Fréchet Inception Distance}
\acrodef{ssim}[SSIM]{Structural Similarity Index Measure}
\acrodef{nisq}[NISQ]{noisy intermediate-scale quantum}
\acrodef{qdm}[QDM]{Quantum Diffusion Model}
\acrodef{crw}[CRW]{Classical Random Walk} % in the text you used random walk (RW) on the first occurrence, sometimes RW and sometimens CRW on the subsequent ones
\acrodef{qc}[QC]{Quantum Computing}
\acrodef{qml}[QML]{Quantum Machine Learning}
\acrodef{qw}[QW]{Quantum Walk}
\acrodef{dtqw}[DTQW]{Discrete-Time Quantum Walk}
\acrodef{roi}[ROI]{Regions Of Interest}
\acrodef{pca}[PCA]{Principal Component Analysis}
\acrodef{kde}[KDE]{Kernel Density Estimation}
\acrodef{ct}[CT]{Computed Tomography}
\acrodef{mri}[MRI]{Magnetic Resonance Imaging}
\acrodef{adam}[ADAM]{Adaptive Moment Estimation}

\section{Introduction}

Recent years have witnessed the rapid ascent of \ac{genai} models, which are \ac{dl} algorithms able to learn the distribution of training data and generate new samples, reflecting specific attributes of the data distribution. Within this field, \acp{dm} have consolidated as one of the top performing models in the state-of-the-art.
Introduced by Sohl et al.~\cite{sohl2015deep} in 2015, \acp{dm} are inspired by the physical process of diffusion, where a system, initially in an ordered state, progressively evolves into a disordered (noisy) state through a sequence of stochastic processes. Their general algorithmic structure is composed of two steps: i) a forward diffusion process in which data samples are gradually corrupted by noise; and ii) a backward denoising process in which a learning algorithm is used to recover the data samples and learn their underlying unknown distribution~\cite{Ho2020}.
The success of \acp{dm} is demonstrated by the variety of applications in which they are used. For example, \acp{dm} are widely used in the healthcare domain for synthetic data generation, image reconstruction and restoration ~\cite{paiano2024transfer}, and to support diagnosis, analysis and prognosis~\cite{Iuliano2024, turrisi2026generating, azad2026systematic}.

In parallel to \ac{ai}, \ac{qc} is progressively gaining attention, as its underlying quantum mechanical principles promise to solve complex computational problems that are resource-intensive or entirely intractable for classical computers.
The intuition behind \ac{qc} relies on the fact that quantum mechanical properties typically require an exponential number of the so-called amplitudes ($\sim2^n$) to describe an $n$-particle system, which for large systems, is beyond the computational capabilities of classical computing.
In contrast, quantum computers require instead a polynomial number of quantum bits (\textit{qubits}) to encode such a system.
Therefore, the promising advantage of \ac{qc} relies on the more efficient encoding and processing of these amplitudes.
Despite the promising potential, \ac{qc} is still in its early stage, and several issues limit its practical applicability. 
In particular, one needs to build such a quantum computer and overcome the unavoidable experimental errors. While quantum error correction procotols are well-known, their full implementation and deployment are still a technological barrier. Nevertheless, we have already quantum computers called \textit{noisy} (i.e. without error correction) that have proved certain advantages over classical computers in specific tasks~\cite{riste2017demonstration,huang2022quantum,Hibat-Allah2024, zhao2026exponential}.

\ac{qml} is a novel research field
which regards the interplay of \ac{ml} techniques and quantum information and technologies. 
From its birth in 2013~\cite{lloyd2013quantumalgorithmssupervisedunsupervised,wittek2014quantumMachineLearning,schuld2015introductionQuantumMachineLearning}, there have been many attempts to reformulate well known classical \ac{ml} problems into their quantum versions. 
However, due to the current limitations of \ac{nisq} devices, today many \ac{qml} algorithms are often restricted to simplified problems, where data are severely reduced in dimensionality, or combined with classical models in the so‑called hybrid quantum–classical frameworks~\cite{McClean2016, Bharti2022}. In recent years, a wide variety of such algorithms has been proposed and investigated, showing the potential, but also the drawbacks, of the interaction between QC and classical ML methods in several domains, such as image classification and pattern recognition~\cite{schuld2014patternclassification, farhi2018, Das2023, Das2024, Geng2022, parigi2025supervised}, chemistry~\cite{Peruzzo2014, Kandala2017}, combinatorial optimization~\cite{Gonzalez2022, venturelli2025investigating}, quantum noise spectroscopy~\cite{martina2023DLenhancedspectroscopy,Martina2022fingerprint,Martina2022fingerprintSoftware,Martina2023noiseClassification}, and sensing~\cite{Haim2025}.

As the interest surrounding \ac{genai} has rapidly increased in recent years, several \ac{qgenai} algorithms have been proposed, including \ac{qgan} and \ac{qvae}.  
A few works showcase applications of these methods for medical image analysis, including data generation~\cite{Khatun_2025}, reconstruction of spine volumes from 2D data~\cite{gadu2023}, or data augmentation for skin cancer detection~\cite{Andra2025,Sudharson26}.

\acp{qdm} are an emerging class of generative algorithms, aimed at running on current quantum computers, that encompass three main approaches, depending on which elements of the \ac{dm} are quantum (the forward process, backward, or both). The theoretical foundations and implementations have been discussed for the first time in our previous work~\cite{parigi2024quantum}. Although in an early stage, \acp{qdm} have been applied in classical domains such as image generation~\cite{cacioppo2023quantumdiffusionmodels,DeFalco2024}, as well as in quantum tasks such as the reconstruction and generation of pure and mixed quantum states~\cite{ZhangQuntao2024, chen2024quantumgenerativediffusionmodel, kwun2025mixed, huang2025continuousqdmStateGenRest, Shah2025, zhu2025channelqdm}. In addition, numerical studies indicate that hybrid quantum-classical designs can enhance the quality of generated data~\cite{DeFalco2024}, learn complex quantum probability distributions~\cite{cacioppo2024quantum2}, reduce the number of trainable parameters required in the reverse denoising process~\cite{DeFalco2024RedPar, zhang2025parameterEfficientQDM}, and overcome dimensional factorization limits in discrete \acp{dm}~\cite{chen2025overcomediscretedm}. 
To the best of our knowledge, only two works have explored the use of \ac{qdm} for medical image analysis~\cite{yeter2025qdmMedImg,chen2025qdmMedMNIST}, both using classical U-Net architecture enhanced with quantum layers for the backward process only. In contrast, 
our method significantly departs form this scheme, as we design a quantum forward and a classical backward process, exploiting the intrinsic noise of current real quantum devices in the forward diffusion process.

This work builds upon our previously proposed   
hybrid quantum-classical \ac{qdm} algorithm~\cite{parigi2025physics}. Not only we present the novel use of the method for medical imaging applications, through studies on several types of datasets, but we also incorporate significant methodological contributions, both in the quantum forward process (e.g. the study of different topologies for the quantum noise process on real devices, and the increment of intensity levels in the Quantum Random Walk component of the method) and the classical backward process (e.g. novel architecture and loss function, with respect to ~\cite{parigi2025physics}).

We apply our method to real-world medical data, in particular grayscale and RGB 2D images and 3D volumes from three datasets (BloodMNIST~\cite{medmnistv2}, BraTS2020~\cite{menze2015bratsbrain, bakas2017tcga, bakas2018brats}, and FractureMNIST3D~\cite{medmnistv2}). Evaluation is performed using three state-of-the-art metrics (\ac{kl} divergence, \ac{fid} and \ac{ssim}). 

Experiments are performed both via simulation of the \ac{qdm} in classical devices, and the experimental implementation on a real quantum computer, employing for the later error mitigation techniques.  
Additionally, we reproduce the main classical \ac{dm} originally proposed in~\cite{austin2021structured, nichol2021improved} in two different setups to perform a comparison between the existing classical methods applied to categorical data and our approach on \ac{nisq} quantum devices. 

This work is organized as follows. The basic principles of \acp{dm} are briefly described in \cref{sec:diffusionModels} and further details focusing on discrete-space \acp{dm} are provided in~\cref{classicalDiffusion} in the supplementary material. The forward process of our hybrid method is described  in~\cref{sec:quantumForward}, focusing on the \ac{dtqw} and its mathematical formulation. Then, we describe the backward model in ~\cref{seq:classicalBack}, and losses and metrics used to validate our experiments in ~\cref{sec:loss}, and ~\cref{sec:metrics}. After describing the datasets used in ~\cref{sec:datasets}, we present the results of the experiments in~\cref{sec:results}, and provide final discussion and conclusions in ~\cref{sec:discussion}.

\section{Methods}

\subsection{Diffusion Models}
\label{sec:diffusionModels}

\acp{dm} are generative algorithms comprising two processes: a \emph{forward} dynamics where data samples are gradually corrupted by noise and a \emph{backward} process where a learning model is trained to reversely recover the data samples~\cite{sohl2015deep, Ho2020}. 
Specifically, given an unknown complex data distribution $q(\mathbf{x}_0)$,
the forward diffusion process is described by a Markov chain that 
progressively transforms an initial data sample 
$\mathbf{x}_0 \sim q(\mathbf{x}_0)$ 
through a sequence of increasingly noisy variables 
$\mathbf{x}_1, \dots, \mathbf{x}_T$ 
into a purely noisy sample 
$\mathbf{x}_T \sim p(\mathbf{x}_T)$, 
where $p(\mathbf{x}_T)$ is called \emph{prior} and is a well-known and tractable distribution such as a Gaussian or a uniform distribution.

More in detail, the forward process is defined through a sequence of conditional 
probabilities of the form $\mathbf{x}_t \sim q(\mathbf{x}_t | \mathbf{x}_{t-1}; \beta_t)$, where 
$\beta_t \in (0,1)$ is a hyperparameter that controls the noise level added at each time step.
The entire forward transition chain is given by:
\begin{equation}
q(\mathbf{x}_{1:T} | \mathbf{x}_0)
= \prod_{t=1}^T q(\mathbf{x}_t | \mathbf{x}_{t-1}; \beta_t),
\label{eq:forward_chain}
\end{equation}
where the factorization follows from the Markov property. 

In the backward denoising phase, the goal is to reverse the forward diffusion 
process. However, because of its stochastic nature, the inverse form cannot be computed 
analytically. Consequently, the flexibility of deep neural networks is exploited to 
approximate the reverse conditional probabilities $p_\theta(\mathbf{x}_{t-1} | \mathbf{x}_t)$ at each time step to 
reconstruct the data sample $\mathbf{x}_{t-1}$ from $\mathbf{x}_t$. The reverse (generative) model therefore is defined by a reverse Markov chain as follows:
\begin{equation}
p_\theta(\mathbf{x}_{0:T})
= p(\mathbf{x}_T)\prod_{t=1}^T p_\theta(\mathbf{x}_{t-1} | \mathbf{x}_t).
\label{eq:reverse_chain}
\end{equation}

The optimization procedure is performed by maximizing the log-likelihood 
$\log p_\theta(\mathbf{x}_0)$ through the variational lower bound:
\begin{equation}
\log p_\theta(\mathbf{x}_0)
\ge 
\mathbb{E}_{q(\mathbf{x}_{1:T} | \mathbf{x}_0)}
\left[
\log p_\theta(\mathbf{x}_{0:T})
- \log q(\mathbf{x}_{1:T} | \mathbf{x}_0)
\right],
\label{eq:loss}
\end{equation}
where $p_\theta(\mathbf{x}_{0:T})$ and $q(\mathbf{x}_{1:T}|\mathbf{x}_0)$ are defined in \cref{eq:reverse_chain} and \cref{eq:forward_chain}, respectively.

In the context of discrete-state space \acp{dm}~\cite{austin2021structured},
data $\mathbf{x}$ are scalar random variables of $K$ categories, where $K \in \mathbb{N}$. 
The conditional probabilities within the forward dynamics are given by 
\begin{equation}
    q(\mathbf{x}_t|\mathbf{x}_{t-1}) = \text{Cat}(\mathbf{x}_t; p = \mathbf{x}_{t-1}\mathbf{Q}_t),
    \label{eq:forwardcategorical}
\end{equation}
where $\text{Cat}(\mathbf{x}; p)$ is a categorical distribution over the one-hot row vector $\mathbf{x}$ with probabilities $p$, and the transition matrix $[\mathbf{Q}_{t}]_{ij} = q(x_t = j \ | \ x_{t-1} = i)$ tells us how likely a walker placed in the state $x_{t-1} = i$ has the chance to move to the site $x_t = j$ after a time-step. 
The prior $p(x_T)$ is a $K-$discrete Uniform distribution. The reverse conditional probability $p_\theta(\mathbf{x}_{t-1} | \mathbf{x}_t)$ has the same functional form of forward transition probability as in \cref{eq:forwardcategorical}.

\begin{figure*}
    \centering
    \includegraphics[width=1\linewidth]
    {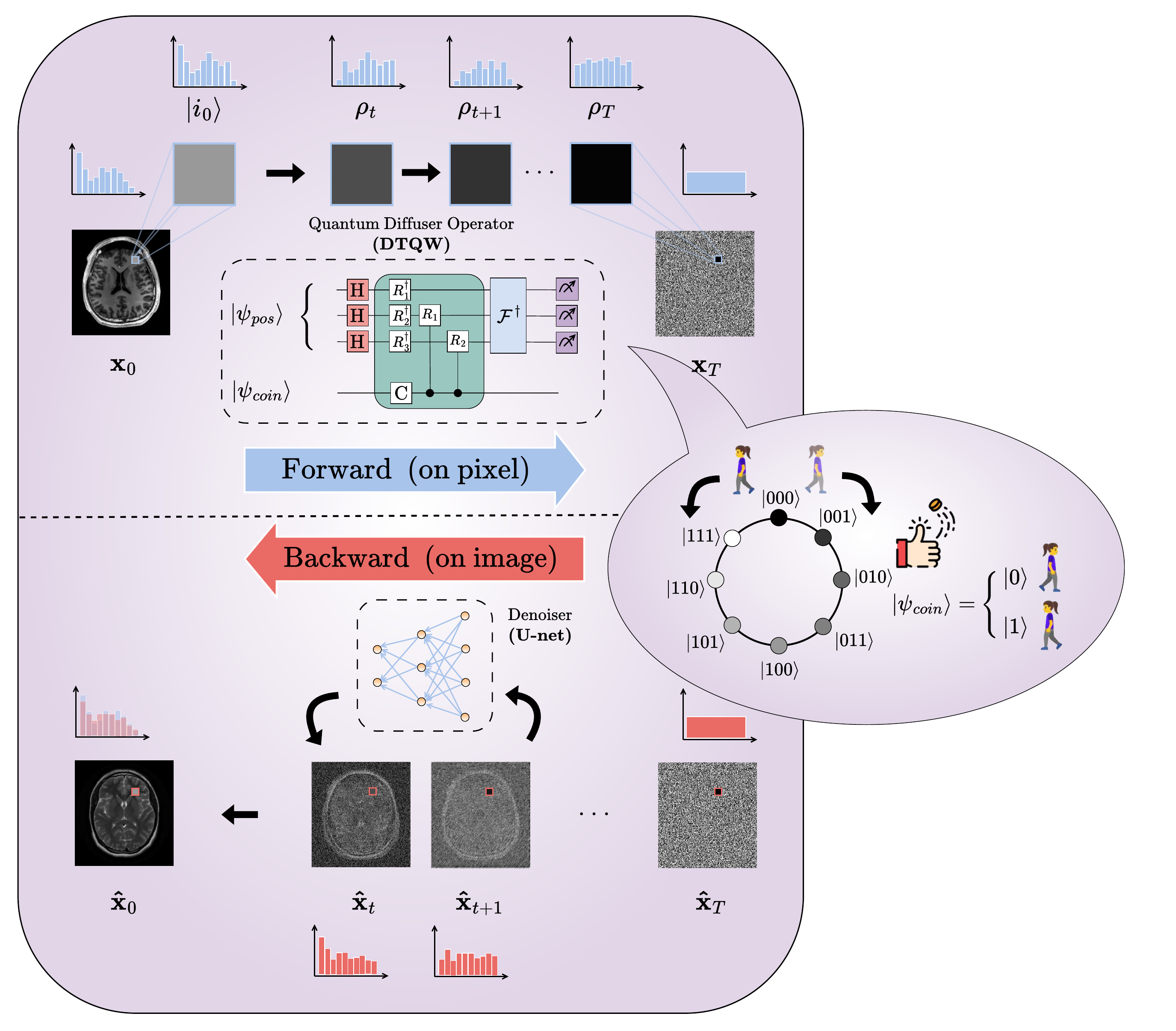}
    \caption{Hybrid quantum diffusion model. Given an initial image $\mathbf{x}_0$ at time $t=0$, every single pixel is associated with a quantum walker whose corresponding color $i_0$ is initially encoded in the quantum state $\ket{i_0}$. The depicted quantum circuit visualizes only one of the multiple time steps of the \ac{dtqw} algorithm, where $\ket{\psi_{pos}}\equiv\ket{i_0}$, $\ket{\psi_{coin}}\equiv\ket{0}$, $H$ are Hadamard gates, $R_k$ are gates $\mathrm{diag}(1,exp(2\pi i/2^k))$ (and their controlled versions $\mathrm{diag}(1,1,1,exp(2\pi i/2^k))$), $\mathcal{F}^\dagger$ is the inverse Quantum Fourier Transform~\cite{Nielsen_Chuang_2010}, and $C$ is the coin operation (implemented in our case with a Hadamard gate). %and $Incr.$, $Decr.$ are the unitaries respectively for the increment and decrement operations.   
    The green-shaded block is repeated for the needed time steps. 
    To obtain the pixel values of $\mathbf{x}_t$ at time $t$ we perform a single measurement of every noisy state $\rho_t$ after $t$ steps of evolution of each walker. Then, for every pixel we reinitialize $\ket{\psi_{pos}}$ with the obtained value, $\ket{\psi_{coin}}$ with $\ket{0}$, and we perform a further single step to obtain the states $\rho_{t+1}$ and the corresponding $\mathbf{x}_{t+1}$. To train the classical backward model, we minimize for all values of $t$ the difference between the pixel-value distribution of $\mathbf{x}_t$ and the output $\mathbf{\hat x}_t$ of a U-Net model that takes $\mathbf{x}_{t+1}$ as input. Finally, to generate new images $\mathbf{\hat x}_0$ we start from samples $\mathbf{\hat x}_T$ of uniform noise and we iteratively apply the trained model.}
    \label{fig:mainfig}
\end{figure*}

\subsection{Quantum forward process}\label{sec:quantumForward}

We deem that the most suitable framework to adapt to currently available \ac{nisq} devices is the one of discrete-state space \acp{dm} as classically treated in~\cite{austin2021structured,nichol2021improved,Hoogeboom2021}. 
In the following, we describe how we implement a Discrete-Time Quantum Walker (\ac{dtqw}) for the forward process of our \ac{qdm}.

As in our precedent work~\cite{parigi2025physics}, we propose a scalable approach where each pixel (or voxel) of the input image is first discretized in values $0,1,\dots,2^{N_q}-1$ and then associated to a \ac{dtqw} executed on a real \ac{nisq} device (for color images, an independent \ac{dtqw} is used for each channel). Transitions between intensity levels define a ring-chain Markov process, as illustrated in the right inset of \cref{fig:mainfig}.

Each node of the ring is a base of the quantum state $\ket{\psi_{pos}}$ and is associated to a possible intensity value $i_0$ for the pixel/voxel. 
In this way, only $N_q$ qubits are necessary to represent $2^{N_q}$ intensity values. An additional qubit $\ket{\psi_{coin}}$ is required for the coin that allows the walker to move right and left (more properly in superposition of right and left) along the chain.

The global state of the system at time 0 is 
\begin{equation}
    \ket{\psi_0}=\ket{\psi_{pos}}\otimes\ket{\psi_{coin}}=\ket{i_0}\otimes\ket{0},
\label{globalState}
\end{equation}
and in density matrix form: $\rho_0=\ket{\psi_0}\bra{\psi_0}$ (See \cref{rho_definition} in supplementary material for more details on quantum formalisms). 

The theoretical dynamics of the walker for $t$ time steps is defined by
\begin{equation}
    \rho_t=U_t\rho_0 U_t^\dagger,
    \label{eq:teoEvolutionQRW}
\end{equation}
where $U_t$ is the unitary corresponding to the circuit shown in the center of \cref{fig:mainfig}, with the green shaded area repeated for $t$ times. This circuit is inspired by the work of ~\cite{razzoli2024efficient}.
However, considering that the circuit is run on a real \ac{nisq} device, with the effects of destructive decoherence phenomena due to the presence of quantum noise, we need to formalize the dynamics adding a quantum channel $\mathcal{E}$ to \cref{eq:teoEvolutionQRW}:
\begin{equation}
    \rho_t=\mathcal{E}(U_t\rho_0 U_t^\dagger).
\end{equation}

The effects of the forward process are probed by single measurements on the qubits encoding different intensity values. In detail, The probability of measuring the walker in the base state $i_t\in[0,2^{N_q}-1]$ is given by
\begin{equation}
    p(i_t)=\mathrm{Tr}(P_i\rho_t P_i)=\mathrm{Tr}(P_i\rho_t),
    \label{eq:probIt}
\end{equation}
with $P_i=\ket{i}\bra{i}\otimes I$ the projector for the $i$-th base ignoring the last qubit for the coin.

The evolution of $\rho_t$ is affected by two different contributing factors: the diffusion induced by the \ac{dtqw} and the intrinsic quantum noise within the machine due to its interactions with the external environment. In fact, a theoretical \ac{dtqw} would never converge to a uniform distribution due to the pure unitary dynamics, which is always invertible. Therefore, the quantum noise is not a detrimental effect but a necessary resource to obtain a final known uniform distribution.

To simplify the practical implementation on \ac{nisq} devices, an equivalent variation of the measurement is proposed, where $i_t$ is classically sampled from
\begin{equation}
    \text{Cat}\left(\mathbf{p}=\left(p(0),p(1),\dots,p(2^{N_q}-1)\right)\right),
    \label{eq:probs}
\end{equation}
where $\text{Cat}(\mathbf{p})$ is a categorical distribution with probabilities defined by $\mathbf{p}$ and $p(0),\dots,p(2^{N_q}-1)$ are the probabilities of \cref{eq:probIt}. In fact, The sampling operation from the categorical distribution is analogous to a quantum measurement of the state, but it allows to decouple the quantum evolution of a generic-pixel \ac{dtqw} from its application to the real images by estimating $p(i)$ with a finite number of shots ($8\,192$ in our case). Moreover,
due to the ring topology on which the \ac{dtqw} evolves, the procedure is further simplified by considering the \ac{dtqw} always starting in $\ket{0}$ and adding the result of the sampling from \cref{eq:probs} to the effective initial position $i_0$ to get to $i_t$.

To train the backward process, we need pairs of values $i_t$ associated with the corresponding $i_{t+1}$. To obtain the latter, we perform a single additional step after \cref{eq:probs} restarting the \ac{dtqw} from $\ket{i_t}$.

The ring-chain topology of our setup is particularly appropriate for \ac{nisq} devices, such as the IBM system employed in our experiments, as the hardware implementation of the qubits in the system naturally maps to such topologies. The \ac{dtqw} is implemented using the Qiskit library from IBM. The forward dynamics is run on the \emph{ibm\_torino} device, which has a total of $133$ qubits with connectivity of $2$ or $3$ qubits.
We select $6$ near qubits (directly connected or with low distance). One of them acts as the coin operator, while the remaining $5$ embed the color intensities. 
In this work, the number of intensity levels is significantly increased (from $8$ to $64$), With respect to our previous work ~\cite{parigi2025physics}. 
In order to obtain this, while maintaining a moderate level of noise in the first time steps, we use the \emph{optimization level} functionality provided by Qiskit (setting it equal to 3). The latter, reduces the number of operation gates during the transpiling phase, i.e. the translation of the circuit to an equivalent one that is compatible with the reduced set of native gates in the IBM quantum devices. 

\subsection{Classical backward process}
\label{seq:classicalBack}
In~\cite{parigi2025physics}, the classical backward model was implemented as a simple multilayer perceptron. 
Here, we propose a more advanced U-Net
model~\cite{ronneberger2015u} and incorporate tailored time embedding mechanism to better capture the diffusion process.

Two versions are proposed. A first, simpler version, that is used for 2D images, adopts the time embedding only on the encoder layers and has a shallow bottleneck. On the other hand, a more complex model, designed for 3D volumes, has the time embedding also in the decoder and is provided with a bottleneck with more layers.

The first model takes as input a batch of 2D data samples of discrete intensity levels
\begin{equation}
    \mathbf{x} \in \{ 0, \ ..., \ 2^{N_q}-1\}^{C \times W \times H},
\end{equation}

corresponding to the positional qubits $N_q$, where $C$, $W$ and $H$ are the number of channels and the image dimensions.
To condition the model on the discrete temporal index $t\in [1,T]$, at each convolutional layer $l$ of the encoder, we sum its input 
$\mathbf{e}^{(l-1)}\in\mathbb{R}^{C_{l-1}\times W_{l-1}\times H_{l-1}}$ 
(with $\mathbf{e}^{(0)}\equiv\mathbf{x}$) to a $C_{l-1}$-dimensional embedding $\mathbf{t}^{(l-1)}$ of $t$ (summing each $\mathbf{t}^{(l-1)}_c$ to the corresponding $W_{l-1}\times H_{l-1}$ matrix $\mathbf{e}^{(l-1)}_c$). 
The structure of the U-Net for 2D data is composed of a sequence of $5$ downsampling encoders consisting of a variable number of convolutional layers which is set by controlling a free parameter $h$. For all the experiments we use $h=1$. Each filter is alternated by a ReLU activation function ($\equiv\max(0,x)$) and a 2D max pooling layer for data compression. 
The other part of the model contains $4$ upsampling decoders that use transpose convolutions for upsampling and are connected to the specular encoder via skip connections.
In detail, the output of the $l$-th upsampling layer over a total of $L=4$, is given by
\begin{equation}
\mathbf{d}^{(l)}
=
\operatorname{cat}\!\left(
\operatorname{up^{(l)}}\!\left(\mathbf{d}^{(l-1)}\right),
\mathbf{e}^{(L-l+1)}
\right),
\end{equation}
with $\mathbf{d}^{(0)}\equiv e^{(L+1)}$, $\operatorname{cat}\!(\cdot,\cdot)$ the concatenation operator and $\operatorname{up^{(l)}}\!(\cdot)$ the $l$-th transpose convolutional layer.

The 3D U-Net instead is composed by $4$ encoding blocks, followed by a longer bottleneck of $3$ convolutional blocks without pooling layers, and $4$ decoding blocks. In this scheme, the $l$-th decoding layer has as input the concatenation of the output of the previous decoding layer $\mathbf{d}^{(l-1)}$ (or the end of the bottleneck for the first) and of the encoding layer $\mathbf{e^{(L-l)}}$. In addition, for this model the time embedding is summed not only to the $\mathbf{e}^{(l)}$ as in the 2D model, but also to the $\mathbf{d}^{(l)}$ for all values of $l$.
Moreover, we adopt group batch normalization for gradient stability and we replace the ReLU activation functions with the SiLU function: $x\cdot\sigma(x)$.

The models are trained using mini-batch gradient descent to minimize the loss described in \cref{sec:loss} between the pixel (or voxel) intensity values of the reference images $\mathbf{x}_t$ and those reconstructed with the U-Net $\mathbf{\hat x}_t$. In detail, at each training iteration, a random $t \in [0, T-1]$ is selected and a batch $\mathbf{x}_0$ of images is independently corrupted for each pixel or voxel with initial value $i_0$ to obtain the corresponding $i_t$ and $i_{t+1}$ as described in \cref{sec:quantumForward}. All those values compose the batches $\mathbf{x}_t$ and $\mathbf{x}_{t+1}$, and the U-Net takes $\mathbf{x}_{t+1}$ as input (conditioned with the value of $t$) and generates $\mathbf{\hat x}_t$. The optimization step is performed using \ac{adam}~\cite{kingma2017adammethodstochasticoptimization} with learning rate equal to $10^{-3}$. We decide to train the 2D U-Nets with images for a fixed number of $20\,000$ epochs. However, the training procedure for the 3D U-Net is longer and the model is susceptible to overfitting owing to its higher complexity. Therefore, for the volumes we decided to reduce the number of epochs to $\sim 4\,000$ by adopting an early-stopping strategy.

After training, the generation of images $\mathbf{\hat x}_0$ is performed by sampling random uniform noise $\mathbf{\hat x}_T$ and iteratively (for $t = T-1,\dots,0$) generating  $\mathbf{\hat x}_t$ using the U-Net with input $\mathbf{\hat x}_{t+1}$ and conditioned with $t$. 

\subsection{Losses}
\label{sec:loss}
We train the backward model using the standard \emph{cross-entropy} loss since it is suitable for categorical data. However, in the case of the BraTS2020 dataset, we decided to introduce a different kind of loss to induce a more accurate pixel-value distribution for the generated data:
\begin{equation}
\begin{aligned}
\mathcal{L}
&=
\sum_{t=1}^{T}
\Biggl(
-\alpha_t \sum_{x\in\mathcal{X}} y_x \log(p_x)
\\
&\qquad
-(1-\alpha_t)
\sum_{x\in\mathcal{X}}
P(x)\log\frac{P(x)}{Q(x)}
\Biggr).
\end{aligned}
\end{equation}
The first term is the cross-entropy, also used for the training of the model with the other two datasets, and the second one is the \ac{kl} divergence between the output and target pixel-value distributions. $\alpha_t\in[0,1]$ is a time-dependent factor that is proportionally scaled with $t$ so that for higher values of $t$ the model learns mainly to match the pixel-value distribution and for lower values the model is trained to reconstruct the image $\mathbf{x}_t$.

\subsection{Metrics}
\label{sec:metrics}
Different metrics are used to evaluate the performance of our model. The \ac{kl} divergence $D_{\text{KL}}(P \ || \ Q)$ computes the divergence between the original pixel‑intensity distribution of the entire dataset and the generated distribution at each time‑step $t$ during forward and backward processes.
Mathematically, \ac{kl} takes the form
\begin{equation}
D_{\text{KL}}(P \ || \ Q) = \sum_{x \in \mathcal{X}} P(x)\log\frac{P(x)}{Q(x)},
\end{equation}
where $P$ and $Q$ are discrete probability distributions over the same support~\cite{kullbackLeibler1951}. \ac{kl} divergence is not symmetric and, conventionally, $P$ is considered the reference distribution and $Q$ the one to compare with, in our case the one of generated images.

The \acf{fid} is a commonly used statistical metric to assess the quality of images generated by models~\cite{heusel2017FID}. It compares the distribution of generated images with the distribution of the set of original images (a \emph{ground truth} set). 
\begin{equation}
\mathrm{FID} = \|\mu_r - \mu_g\|_2^2 + \text{tr} \left(\Sigma_r + \Sigma_g - 2(\Sigma_r \Sigma_g)^{1/2}\right)
\end{equation}
Here, $\mu_r$ and $\mu_g$ represent the mean feature vectors of real and generated samples extracted using a pre-trained \emph{Inception-v3} network, while $\Sigma_r$ and $\Sigma_g$ denote their corresponding covariance matrices.
A lower value of \ac{fid} indicates better generative performance, whereas a higher value reflects poorer performance. 

We also calculate the \acf{ssim}~\cite{wang2003msssim}
between the generated image and the real reference image to assess perceptual quality,
with an emphasis on structural information. Indeed, \ac{ssim} evaluates image similarity by
incorporating perceptual aspects of the human visual system and produces an index in the
range $[-1, 1]$, where higher values indicate greater structural fidelity. Mathematically, the \ac{ssim} between two images $x$ and $y$ is defined as
\begin{equation}
\text{SSIM}(x, y) =
\frac{(2\mu_x\mu_y + c_1)(2\sigma_{xy} + c_2)}
{(\mu_x^2 + \mu_y^2 + c_1)(\sigma_x^2 + \sigma_y^2 + c_2)},
\end{equation}
where $ \mu_x $ and $ \mu_y $ denote the mean pixel intensities of the two image distributions, $ \sigma_x $ and $ \sigma_y $ their standard deviations, and $ \sigma_{xy} $ the covariance. The constants $ c_1 = (k_1 L)^2 $ and $ c_2 = (k_2 L)^2 $ stabilize the division, where $ L $ is the dynamic range of the pixel values (typically $ 2^{\text{bits per pixel}} - 1 $), and the default parameters are $ k_1 = 0.01 $ and $ k_2 = 0.03 $.

Finally, to visualize, compare, and quantify the original and generate data distributions, we perform a \ac{pca} on the datasets to identify the two components corresponding to the highest eigenvalues. We then project the corresponding samples onto a 2D plane and apply a \ac{kde}
to evaluate the degree of overlap between the two distributions.
\ac{kde} is given by the following
\begin{equation}
    \hat{f}_h(x) = \frac{1}{n}\sum_{i=1}^n K_h (x-x_i), 
\end{equation}
where $K$ is a kernel function and $h > 0$ is a smoothing width parameter which determines the density approximation.

\subsection{Medical datasets}\label{sec:datasets}
To test our algorithm, we employ three distinct datasets of different nature, complexity and dimension. Firstly, we consider a dataset of $3\,329$ images of $64\times 64$ RGB color pixels extracted from the class 6 of BloodMNIST (where training, validation and test sets are merged together) included in the MedMNIST collection of datasets~\cite{medmnistv2}. Each channel of the color images is quantized with values in $[0,31]$. Next, we use a real dataset composed of  multi-modal brain \ac{mri} scans (comprising tumors, malformations, etc.) obtained from the BraTS2020 dataset~\cite{menze2015bratsbrain, bakas2017tcga, bakas2018brats}. The volumes that compose the original dataset are preprocessed extracting a single slice in the $z$ direction and a single mode to obtain $484$ grayscale images cropped to $190 \times 190$ pixels and quantized in the range $[0,63]$.
Lastly, we adopt the 3D dataset FractureMNIST3D~\cite{medmnistv2} also from MedMNIST, which contains $1\,370$ grayscale volumes of size $28 \times 28 \times 28$ of rib fractures from \ac{ct} scans. We quantize the voxel values in the range $[0,31]$. 

\section{Results}
\label{sec:results}

To train the models on our datasets, we first gather forward probabilities running the \ac{dtqw} on real IBM \ac{nisq} devices starting in $\ket{0}$ as described in \cref{sec:quantumForward}. Then,
for each one of the datasets, composed of images $\textbf{x}_0$, we adapt the forward process to the real data, reproducing the dynamics of independent \acp{dtqw} on each pixel, obtaining $\mathbf{x}_t$ for $t=1,\dots,T$. Finally, we train the model to reconstruct the images $\mathbf{\hat{x}}_t \approx \mathbf{x}_t$ at time step $t$ based on the forward processes from $\mathbf{x}_t$ to $\mathbf{x}_{t+1}$ as described in \cref{seq:classicalBack}. The generation of new samples $\mathbf{\hat x}_0$ is performed after training, starting from pure noise $\mathbf{\hat x}_T$ and iteratively generating $\mathbf{\hat{x}}_{T-1},\mathbf{\hat{x}}_{T-2},\dots,\mathbf{\hat{x}}_{0}$ using $\mathbf{\hat x}_{t+1}$ as input to generate $\mathbf{\hat x}_t$.
The data pre-processing, the training of the model and its backward process are realized with the PyTorch library.

\begin{table*}[t!]
\centering
\caption{Quantitative evaluation across the three datasets using \ac{kl} divergence, \ac{fid}, and \ac{ssim}. Lower \ac{kl} divergence and \ac{fid} indicate better distributional alignment, while higher \ac{ssim} means good structural similarity. 
The metrics are calculated for $100$ generated samples and, regarding the 3D volumes, the reported \ac{fid} value corresponds to the average of different volume slices ($z=5, 10, 15, 20, 25$).
}
\renewcommand{\arraystretch}{1.3}
\setlength{\tabcolsep}{7.15pt}

\begin{tabular}{|p{2.5cm}||
                c|c|c||
                c|c|c||
                c|c|c|}
\hline
\multicolumn{10}{|c|}{\textbf{Model evaluation on 2D and 3D images}} \\
\hline

\multirow{3}{*}{\textbf{Dataset}}
& \multicolumn{3}{c||}{\textbf{KL}}
& \multicolumn{3}{c||}{\textbf{FID}}
& \multicolumn{3}{c|}{\textbf{SSIM}} \\
\cline{2-10}

& \multicolumn{2}{c|}{Classical}
& \multirow{2}{*}{Quantum}
& \multicolumn{2}{c|}{Classical}
& \multirow{2}{*}{Quantum}
& \multicolumn{2}{c|}{Classical}
& \multirow{2}{*}{Quantum} \\
\cline{2-3}\cline{5-6}\cline{8-9}

& $\beta=0.9$ & $\beta=\beta_t$
&
& $\beta=0.9$ & $\beta=\beta_t$
&
& $\beta=0.9$ & $\beta=\beta_t$
& \\
\hline

BloodMNIST
& 0.098 & \textbf{0.046}
& 0.090
& 274.4 & \textbf{166.6}
& 243.3
& 0.220 & 0.336
& \textbf{0.377} \\
\hline

BraTS2020
& 0.396 & 0.192
& \textbf{0.124}
& 280.3 & \textbf{266.8}
& 297.7
& 0.646 & 0.569
& \textbf{0.676} \\
\hline

FractureMNIST3D
& 0.047 & 0.013
& \textbf{0.011}
& 169.6 & 100.1
& \textbf{89.70}
& \textbf{0.564} & 0.550
& \textbf{0.564} \\
\hline

\end{tabular}
\label{tab:metrics}
\end{table*}

In \cref{tab:metrics} we report the evaluation of \ac{kl} divergence, \ac{fid} and \ac{ssim} comparing the original BloodMNIST and BraTS2020
datasets. 
The different \ac{ssim} values obtained for the BloodMNIST and BraTS2020 datasets can be explained by the distinct properties of the corresponding images: BloodMNIST images have a lower resolution (sizes of the images) and number of intensity levels than the BraTS2020 brain images. Thus, lower spatial resolution and stronger quantization degrade structural details, luminance, and contrast, directly impacting the three components on which \ac{ssim} is computed. We use $100$ samples to evaluate all the metrics we reported in the \cref{tab:metrics}. This means that, once the training and the generation phases are finished, we select a batch of $100$ random images from both the datasets of original and generated images/volumes and we compute the \ac{kl} divergence, \ac{fid} and \ac{ssim} respectively. 

Overall, we can observe that our quantum hybrid model is competitive with respect to its classical counterpart. Especially with the complex 3D data of FractureMNIST3D, we obtain data with lower \ac{kl} divergence and \ac{fid} and equal \ac{ssim}. Also, for all three datasets we obtain higher values of \ac{ssim} for the quantum model respect to the classical one. This denotes that the quantum forward may help to obtain better structural features in the generated data. Regarding the poor results for quantum models considering the \ac{fid} on BloodMNIST and BraTS2020, we need to keep in mind that \ac{fid} is calculated using a model pretrained on generic high-quality data. This means that the \ac{fid} can be a poor choice for a metric when using specialized data as in our case.

In the next sections we focus on the results regarding the three considered datasets. Further generated samples are shown in supplementary material.

\subsection{Quantum diffusion on RGB BloodMNIST images}

The first dataset that we consider is composed of relatively small $64\times64$ color images of $32$ pixel-intensity levels for each RGB channel.
We use $6$ connected qubits, $1$ acting as the coin operator, while the remaining $5$ embed the color intensities. 

\begin{figure}[t!]
    \centering
    \subfloat[Original\label{fig:imageBloodComparisonOrig}]{%
        \includegraphics[width=\linewidth]{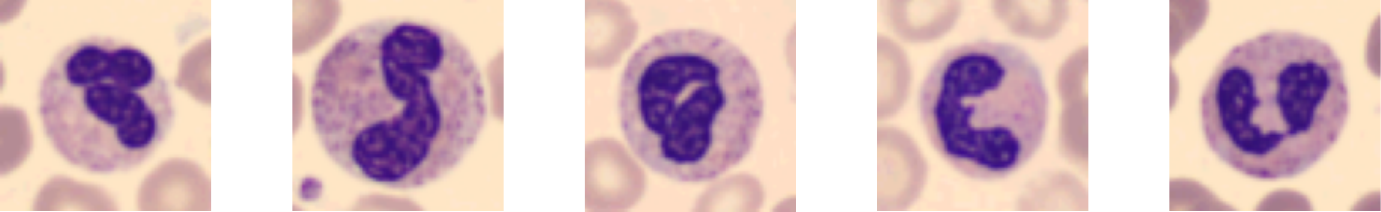}%
    }
    \hfill
    \subfloat[Generated\label{fig:imageBloodComparisonGen}]{%
        \includegraphics[width=\linewidth]{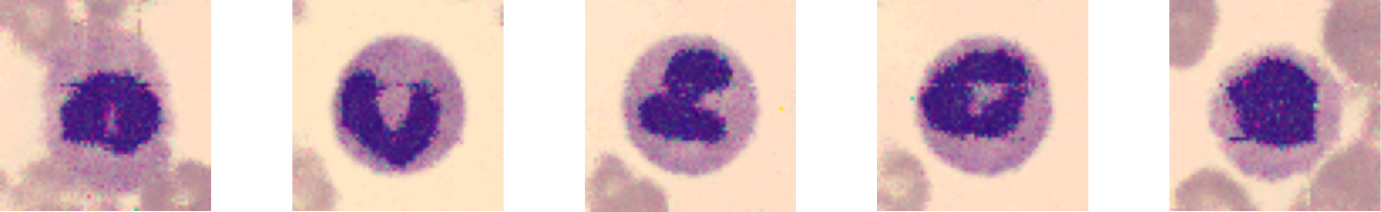}%
    }
    \caption{Five samples of the original set of BloodMNIST images    Fig.~\ref{fig:imageBloodComparisonOrig} and five samples of images generated by our model Fig.~\ref{fig:imageBloodComparisonGen}. Each image has a shape of $64\times64$ with $32$ intensity levels for each RGB channel.}
    \label{fig:imageBloodComparison}
\end{figure}

\begin{figure}[ht!]
\includegraphics[width=0.4\textwidth]{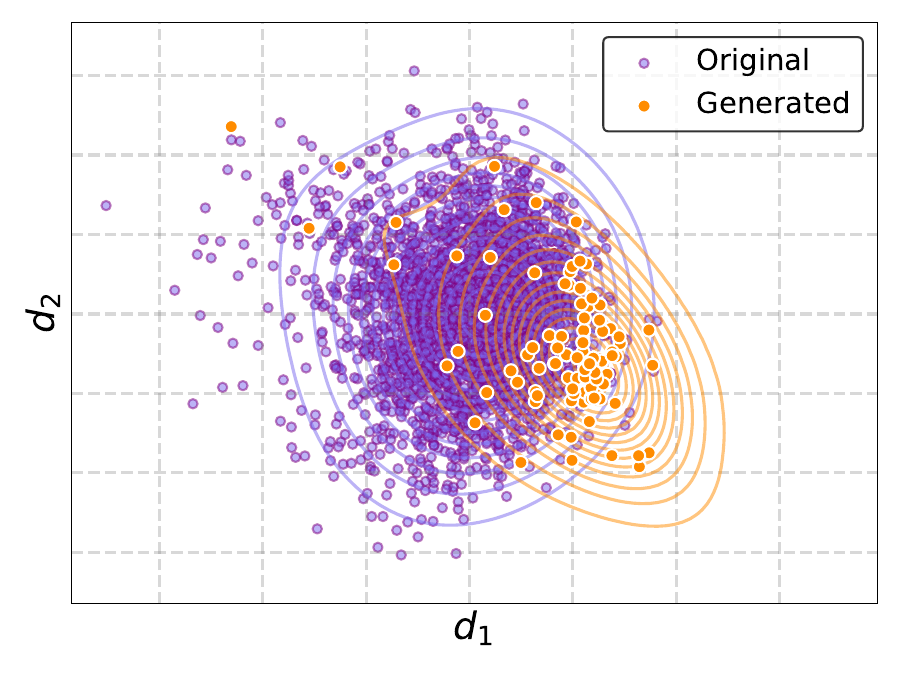}
    \caption{BloodMNIST distribution of the two largest eigenvalues of the \ac{pca} for $3\,000$ original samples (purple) and 100 generated samples (orange). The contours represent the \ac{kde}.}
    \label{fig:imageBloodPCA}

\end{figure}

The trained model is capable of generating images reflecting the original composition. As visible in \cref{fig:imageBloodComparison}, the generated images possess all the main features of the original images, i.e., the centered blood cell with its inner corpuscles and the surrounding cells. In \cref{fig:imageBlood} in the supplementary material, we also show more details on the forward and backward processes after the training. The qualitative visual analysis is also supported by the quantitative evaluation reported in \cref{tab:metrics}. The main complexity of training our hybrid model with the BloodMNIST dataset, is the fact that the images are in color and our quantum forward process is designed to operate independently for each channel (and pixel). This is an evidence of the adaptability of our \ac{qdm} approach to multi-channel data.

To assess the generative capabilities of the model and how it is capable to cover the original dataset variance, we show in \cref{fig:imageBloodPCA} the \ac{pca} in two dimensions of the entire BloodMNIST dataset compared to 100 generated samples. Our model is capable of learning the two modes of the dataset distribution and generating samples in both of them.

\subsection{Quantum diffusion on BraTS2020 images}
To evaluate the applicability of our model to different and more complex data, we decided to adopt a dataset of bigger $190\times190$ grayscale images of $64$ pixel-intensity levels.
We use $7$ connected qubits, $1$ acting as the coin operator, while the remaining $6$ embed the color intensities. 

\begin{figure}[ht!]
    \centering
    \subfloat[Original\label{fig:imageBrainComparisonOrig}]{%
         \includegraphics[width=\linewidth]{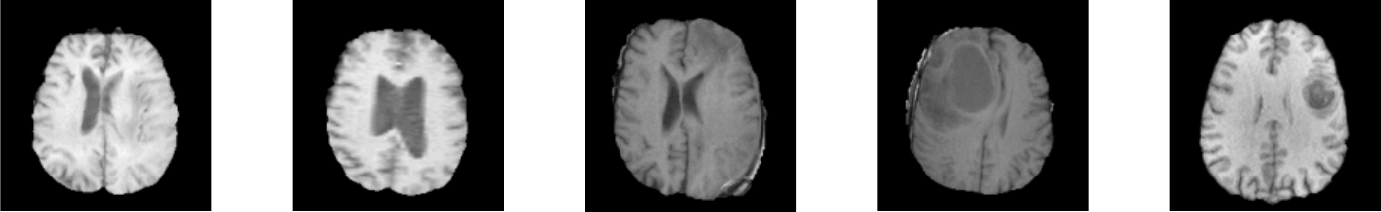}}
    \hfill
    \subfloat[Generated\label{fig:imageBrainComparisonGen}]{%
         \includegraphics[width=\linewidth]{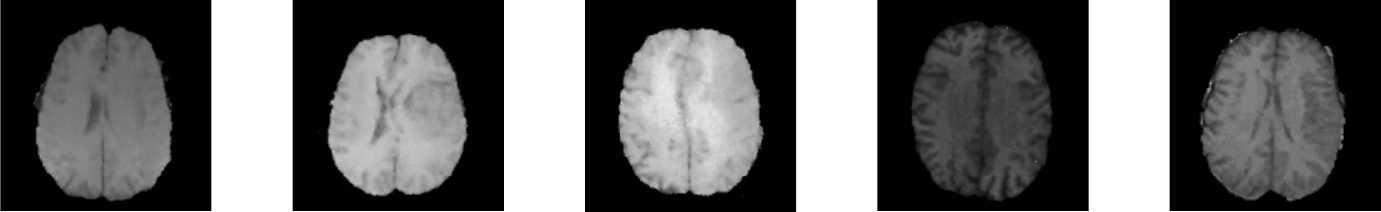}}
    \caption{Five samples of the original set of BraTS2020 images Fig.~\ref{fig:imageBrainComparisonOrig} and five samples of the brain images generated by our model Fig.~\ref{fig:imageBrainComparisonGen}. Each image has a shape of $190\times190$ with $64$ intensity levels.}
    \label{fig:imageBrainComparison}
\end{figure}

\begin{figure}[ht!]
\includegraphics[width=0.4\textwidth]{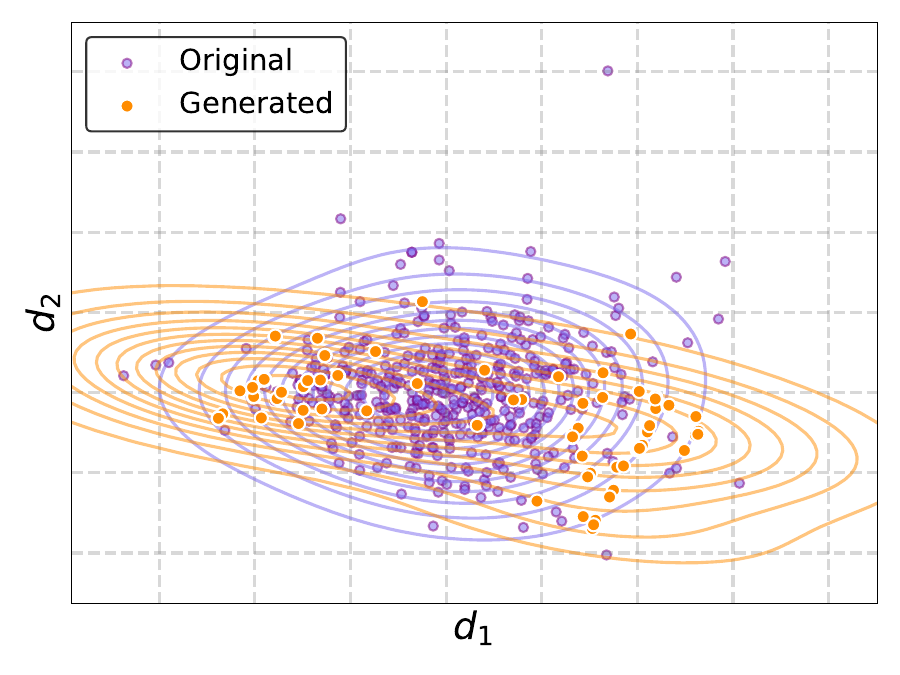}
    \caption{BraTS2020 images distribution of the two largest eigenvalues of the \ac{pca} for 450 original samples (purple) and 100 generated samples (orange). The contours represent the \ac{kde}.}
    \label{fig:imageBrainPCA}
\end{figure}

As visible in the examples of \cref{fig:imageBrainComparison}, the generative model is capable of reproducing the complex features of the brain slices from BraTS2020. In particular, apart from the correct generation of the brain shapes, the generated images show different levels of intensity, as the real ones, and different shapes for the ventricles and folds. Moreover, the model was also able to generate the tumoral masses in some of the samples.

The data are also generated with a variance comparable to the real dataset, as visible in the \ac{pca} of \cref{fig:imageBrainPCA}. Contrarily to the images of BloodMNIST, the \ac{pca} of BraTS2020 exhibits a single mode. Nevertheless, the generative model does not suffer from model collapse.

\subsection{Quantum diffusion on FractureMNIST3D volumes} \label{sec:fracture}
Finally, we focus on a 3D dataset. In this case we use $28\times 28\times 28$ grayscale volumes with 32 intensity levels. Like for the BloodMNIST dataset, we use 6 qubits for the forward process.

\begin{figure}[ht!]
    \centering
    \subfloat[Original\label{fig:VolRenderingOrig}]{%
         \includegraphics[width=1\linewidth]{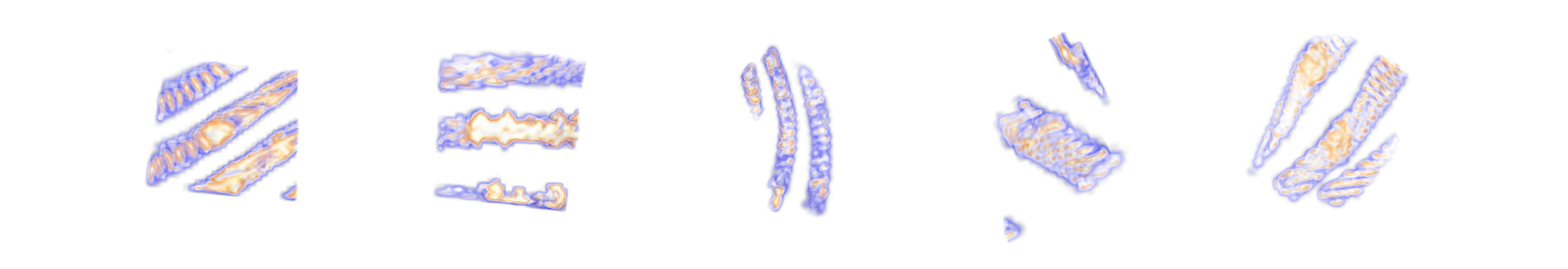}}
    \hfill
    \subfloat[Generated\label{fig:VolRenderingGenGood}]{%
         \includegraphics[width=1\linewidth]{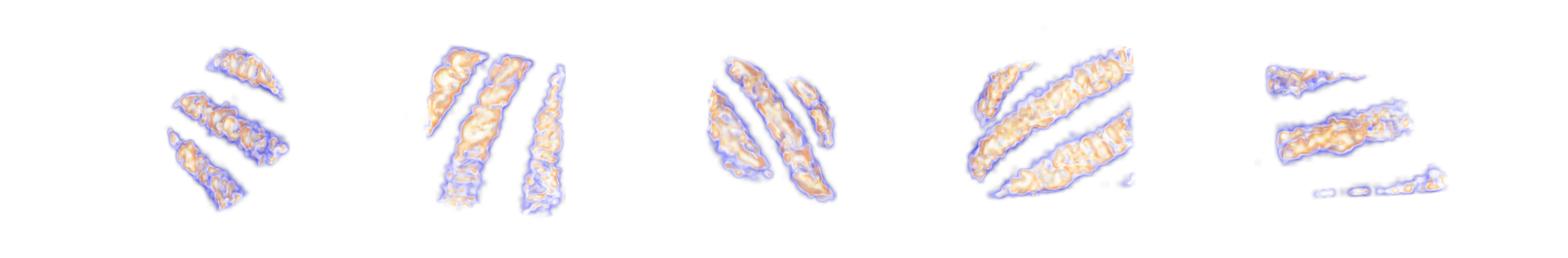}}
    \caption{Volume rendering of $5$ distinct samples from Fig.~\ref{fig:VolRenderingOrig} the original dataset FractureMNIST3D and Fig.~\ref{fig:VolRenderingGenGood} generated volumes produced by the 3D U-Net. Renderings generated using 3D Slicer.
    }
    \label{fig:VolRendering}
\end{figure}

\begin{figure}[ht!]
    \centering
    \includegraphics[width=0.385\textwidth]{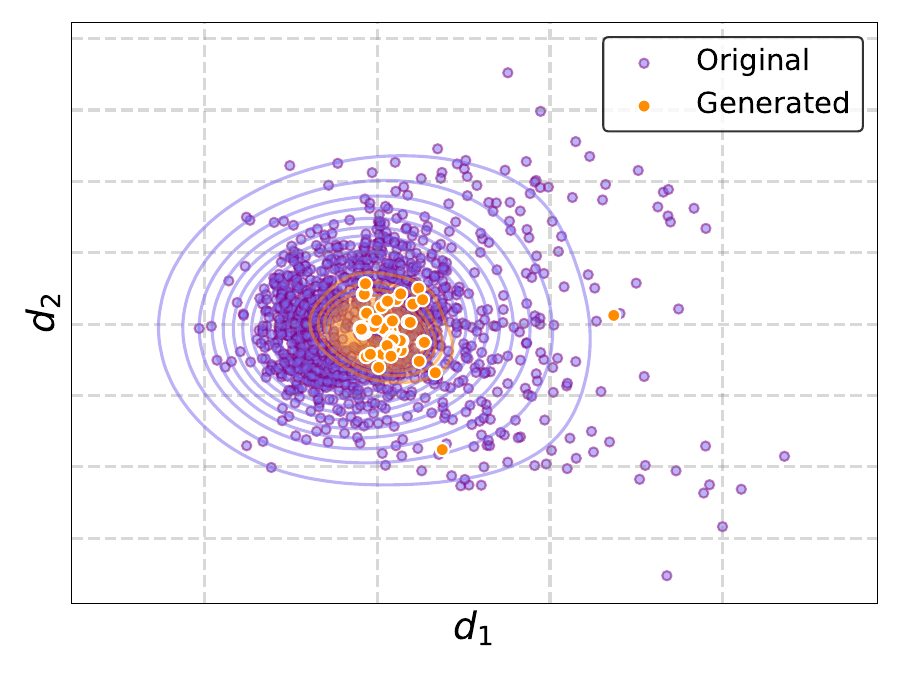}
    \caption{FractureMNIST3D distribution of the two largest eigenvalues of the \ac{pca} for 1370 original samples (purple) and 100 randomly generated samples (orange). The contours represent the \ac{kde}.}
    \label{fig:pcaFracture}
\end{figure}

This setting shows that our approach can also be adapted to data with more than two dimensions. As detailed in \cref{seq:classicalBack}, the backward process is modified for 3D volumes using a 3D U-Net instead of the common 2D U-Net used for the other datasets. The trained model is capable to generate volumes that present the same characteristics of the original data, as visible in the second row of \cref{fig:VolRendering}. 
These figures 
are obtained by applying a segmentation editor using a fixed automatic threshold using 3D Slicer \cite{Fedorov2012}, and by performing a customized volume rendering to highlight the voxel distribution. With this set up we are able to obtain consistent volumes, despite some samples are not correctly generated as we expect. 
Moreover, we calculate the quantitative metrics of the last rows of \cref{tab:metrics}. The values of \ac{kl} divergence and \ac{ssim} are calculated as for the previous datasets. Instead, the calculation of the \ac{fid} uses features extracted by an Inception V3 model pretrained on images and not volumes. For this reason, we decided to calculate the \ac{fid} value for the different slices $z=5,10,15,20,25$ and report the average value of them.
In \cref{fig:pcaFracture}, we show the \ac{pca} in two dimensions of the original and generated data.

\section{Discussion and conclusions}\label{sec:discussion}

The state-of-the-art of hybrid and quantum algorithms applied to real-world data often faces challenges due to the impossibility for current \ac{nisq} devices to deal with large and resource-intensive data such as high-resolution images and volumes.
Our work contributes to different fronts and circumvent this issue: first, we demonstrate that current available \ac{nisq} devices can be used to model the forward process of a \ac{dm}, without the need of error correcting codes, directly exploiting intrinsic quantum noise of the machine. In fact, quantum noise is a constituent factor that is necessary for the convergence to the uniform distribution in the forward process of our model. 
Second, we demonstrate that our hybrid quantum forward dynamics can handle sophisticated and complex data samples such as RGB colored images,
large-scale grayscale images, and 3D volumes. 

We mentioned in \cref{sec:quantumForward} that an important aspect of our architecture is the quantum noise during the forward diffusion dynamics. In our work, in fact, we have the dual effect of the evolution of the \ac{dtqw} and the natural degradation of the quantum state with time. There is evidence that a (moderate) contribution of quantum noise to a \ac{qw} dynamics helps to accelerate the diffusion process~\cite{DallaPozza2022reinforcement}. At the same time, quantum noise can influence the measurement results~\cite{Martina2022fingerprint,Martina2022fingerprintSoftware} in ways that are more difficult to discriminate with the increase in noise complexity~\cite{Martina2023noiseClassification}. The quantum noise factor in our work could behave as an accelerator of the diffusion dynamics in ways that are either compatible with the generative model and at the same time more complex than classical diffusion processes.

Another important aspect of our work relies in the current \ac{dtqw} algorithm used to perform the diffusion process on a quantum device. Such algorithm represents an efficient method to converge to a target distribution after a moderate number of time-steps, by taking into account the topology of the IBM machine. In fact, a necessary condition for the algorithm to work properly is to select the coin register with 3 degree connections, as it can drive the diffusion of the \ac{qw} to proceed forward or backward \cite{razzoli2024efficient}.
Using a cyclic topology allows us to execute a single quantum walker forward dynamics for a tunable number of time-steps and to add the results, in terms of probabilities of occupying certain basis states at each $t$, to the remaining pixels and to sample the associated image distribution. 
Nevertheless, the specific \ac{dtqw} can be used for additional topologies. 

Lastly, we want to emphasize the fact that despite the presence of a classical neural network algorithm to reversely denoise the image, the forward dynamics originate from a quantum process and use a real quantum device with quantum noise. Thus, although we access to classical quantities (probabilities) when we perform the measurement, the process that produces such quantities is entirely quantum, and we cannot have a direct access to it efficiently by quantum mechanical postulates. On the contrary, our backward model is sufficient to learn how the unknown noise samples act on the pixel values and to appropriately reconstruct the backward dynamics.

The application of \acp{qdm} to medical images is still in its early stage.
Even if our model has a different architecture and is trained on different medical datasets with respect to the other works, the results we obtain are comparable. Although our \ac{fid} value is slightly higher than~\cite{yeter2025qdmMedImg, chen2025qdmMedMNIST}, we obtain comparable or higher values for the \ac{ssim} considering our models trained on BraTS2020 brain images and FractureMNIST3D volumes. This means that our model can capture the entire structure of the images.
However, the lower \ac{ssim} values observed in our generated images for BloodMNIST compared to those reported in~\cite{yeter2025qdmMedImg} can be justified as follows. In our setting, the images are limited to a resolution of $64\times 64$ pixels for BloodMNIST with only 32 color intensity levels. Whereas~\cite{yeter2025qdmMedImg} employs images of $1024\times 1024$ pixels with $256$ intensity levels. The combination of low spatial resolution and stronger quantization in our case severely degrades structural details, luminance, and contrast, directly affecting the three components on which \ac{ssim} is calculated~\cite{Wang2004ErrorVisibilitySSIM}. In contrast, larger images with higher intensity depth preserve structural characteristics much more effectively, resulting in higher and more stable \ac{ssim} values. Moreover, the nature of the images in our work is completely different with respect to that of other works.

This is a proof of concept that shows the applicability of available quantum computers in the field of healthcare. With this work, we would like to contribute to the research and development of medical applications using quantum computers as well as testing the capabilities of current \ac{nisq} devices on such critical and highly demanding problems.

\section*{Code Availability}
The source code that was used for the design and training of the models is available in the following GitHub repository: \url{https://github.com/trianam/quantumDiffusionModelsMedicalImageAnalysis} .

\section*{Acknowledgments}
M.P.,~S.M.~and~F.C. acknowledge financial support from the PNRR MUR project PE0000023-NQSTI. S.M~and~F.C. also acknowledge financial support from Fondazione Cassa di Risparmio di Firenze.
A.C.-L. acknowledges funding from Grant RYC2022-037769-I funded by MICIU/AEI/10.13039/501100011033 and by “ESF+". F.A.V. and M.A.G.B. acknowledge funding from the Maria de Maeztu Units of Excellence Programme CEX2021-001195-M, funded by MICIU/AEI/10.13039/501100011033, as well as the European Union under the ERC Synergy Grant Zee-Zoom-Zap (grant no. 101224844).

\bibliographystyle{unsrt}
\bibliography{biblio}

\pagebreak
\widetext
\pagebreak

\setcounter{equation}{0}
\setcounter{figure}{0}
\setcounter{table}{0}
\setcounter{page}{1}
\makeatletter

\renewcommand{\thesection}{S}
\renewcommand{\theequation}{S\arabic{equation}}
\renewcommand{\thefigure}{S\arabic{figure}}

\section*{Supplementary material}

\subsection{Quantum computing}
\label{ref:qc}

Quantum computing is a computational paradigm that exploits quantum physical phenomena to encode and process information~\cite{Nielsen_Chuang_2010}. Its building blocks are called \emph{qubits} (quantum bits), in analogy to the classical bits. A single-qubit state $\ket{\psi}$ can store and represent bits in quantum states labeled as $\ket{0}$ and $\ket{1}$, but it can also generate a superposition between these two, namely $\ket{\psi} = \alpha\ket{0}+\beta\ket{1}$, with $\alpha,\beta\in\mathbb{C}$, referred to as \emph{probability amplitudes} that satisfy $|\alpha|^2+|\beta|^2=1$. Multiple qubits can hold an exponential number of bit-strings in those superpositions, i.e. an $n-$qubit state can contain a superposition of all possible $2^{n}$ bit-strings $\ket{\psi}=\sum_{i=1}^{n}c_{i}\ket{e_{i}}$, where $\{\ket{e_{i}}\}$ is a set of all possible basis states, and $c_{i}\in\mathbb{C}$ with $\sum_{i=1}^{n}|c_{i}|^2=1$. Notice that $n$ information blocks (qubits) can store an exponential amount of classical information (bit-strings). However, these superpositions are not physical observables. At the final stage of the quantum computation, the qubit state $\ket{\psi}$ is \emph{measured}, obtaining a classical bit-string $e_i$ of information with probability $|c_i|^2$, and the qubit-state \emph{collapses} into the quantum state $\ket{e_i}$. Repeating the exact computation will lead to a probability distribution of all the bit-strings with probability of each of them equal to $|c_{i}|^2$. In other words, quantum computing is a probabilistic model of computation that requires to repeat the same operations a certain number of times to obtain the probability distribution of the bit-strings to be processed later. Therefore, quantum algorithms design a sequence of operations to manipulate these qubit-states in a way that with only few measurements one can reconstruct a meaningful probability distribution that encodes the solution of the problem at hand.

In the quantum circuit model, computation is expressed as a sequence of quantum gates, each represented by a unitary operator acting on one or more qubits. Mathematically speaking, qubit states are vectors in a $2^n$ dimensional Hilbert space.
For instance, the single-qubit state can be represented by a column vector \begin{equation}
    \ket{\psi} =
\begin{bmatrix}
\alpha \\ \beta
\end{bmatrix} \quad \text{where} \quad 
\ket{0} =
\begin{bmatrix}
1 \\ 0
\end{bmatrix},
\
\ket{1} =
\begin{bmatrix}
0 \\ 1
\end{bmatrix}.
\end{equation}

Operations on qubit states are implemented via \emph{quantum logic gates}, which correspond to unitary transformations represented by matrices $U$ of dimension $2^m\times 2^m$, where $m$ is the number of qubits that they act on, satisfying the condition $U^\dagger U = I$, where $\dagger$ denotes the conjugate transpose. Quantum gates can be sequentially combined to form quantum circuits, enabling the manipulation of quantum information. In the case of a single qubit, these transformations are described by $2\times 2$ unitary matrices. A characteristic example is the Hadamard gate that transforms quantum basis states into a superposition state:
\begin{equation}
H =
\begin{bmatrix}
\frac{1}{\sqrt{2}} & \frac{1}{\sqrt{2}} \\
\frac{1}{\sqrt{2}} & -\frac{1}{\sqrt{2}}
\end{bmatrix},
\quad
H|x\rangle = \frac{|0\rangle + (-1)^x |1\rangle}{\sqrt{2}}, \quad x \in \{0,1\}.
\end{equation}
A relevant example of a two-qubit gate is the CNOT gate, which enables the creation of entangled states as
\begin{equation}
\mathrm{CNOT} =
\begin{bmatrix}
1 & 0 & 0 & 0 \\
0 & 1 & 0 & 0 \\
0 & 0 & 0 & 1 \\
0 & 0 & 1 & 0
\end{bmatrix}.
\end{equation}
The action of the CNOT gate on the four basis states is:
\begin{equation}
\begin{aligned}
&\ket{00}\mapsto \ket{00}, \quad \ket{01}\mapsto\ket{01}, \\
&\ket{10}\mapsto \ket{11}, \quad\ket{11} \mapsto \ket{10}.\\
\end{aligned}
\end{equation}

In practice, creating and holding the qubit superposition is hard. The required quantum phenomena are extremely fragile, which implies that errors generate and accumulate during the computation. This problem can be circumvented with the use of quantum error correcting codes and fault-tolerant quantum operations, but state of the art quantum devices are not capable of such implementations yet. Therefore, we are now in the so-called \ac{nisq} era \cite{Preskill2018}, where devices can only hold a number of qubits in the range of tens to hundreds, and we have to coexist with quantum noise and errors in our computation. Thus, hybrid quantum-classical algorithms, such as Variational Quantum Algorithms \cite{Bharti2022}, have emerged as potential short and mid-term solutions to apply quantum computing to real-world scenarios. These algorithms contain a classical optimization subroutine that adjusts the parameters of the quantum circuit to balance the noise coming from the quantum computer. There also exist quantum error mitigation techniques \cite{Cai2023} that help in reducing the errors from the quantum devices or provide better noiseless estimations of the results of the quantum computation.

Consequently, to describe realistic scenarios of quantum computation, and for a better understanding of our work, it is useful to adopt the density-operator formalism.
In this framework, the quantum state is represented by a $2^n \times 2^n$ matrix $\rho$ that satisfies the following properties:  
\begin{equation}
\rho = \rho^{\dagger}, \qquad 
\rho \ge 0, \qquad 
\mathrm{Tr}(\rho)=1.
\label{rho_definition}
\end{equation} 
In particular, the first ensures that its eigenvalues are real quantities, the second avoids negative probabilities, and the last one ensures that the probabilities sum to $1$.
For a single-qubit described by $\ket{\psi}$, the corresponding density matrix takes the form:
\begin{equation}
\rho = |\psi\rangle \langle\psi| = 
\begin{pmatrix}
|\alpha|^2 & \alpha\beta^* \\
\alpha^*\beta & |\beta|^2
\end{pmatrix}.
\end{equation}
Here, the diagonal elements represent the probability of measuring the system in the states $\ket{0}$ and $\ket{1}$, while the off-diagonal elements encode the quantum \emph{coherence} of the system, which is responsible for interference phenomena.

\subsection{Classical random walks and discrete-state space diffusion models}
\label{classicalDiffusion}
\begin{figure}[ht]
    \centering
    \subfloat[First-neighbor Markov chain.\label{fig:first-neighbor-markov}]{%
        \includegraphics[width=0.45\linewidth]{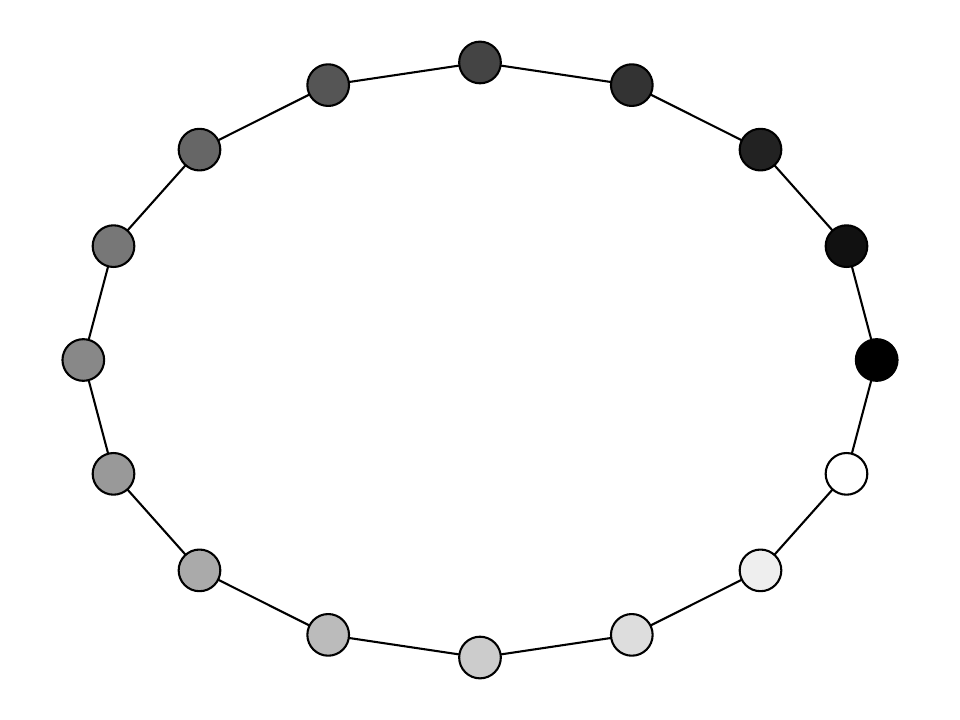}}
    \hfill
   \subfloat[Fully connected Markov chain.\label{fig:full-markov}]{%
        \includegraphics[width=0.45\linewidth]{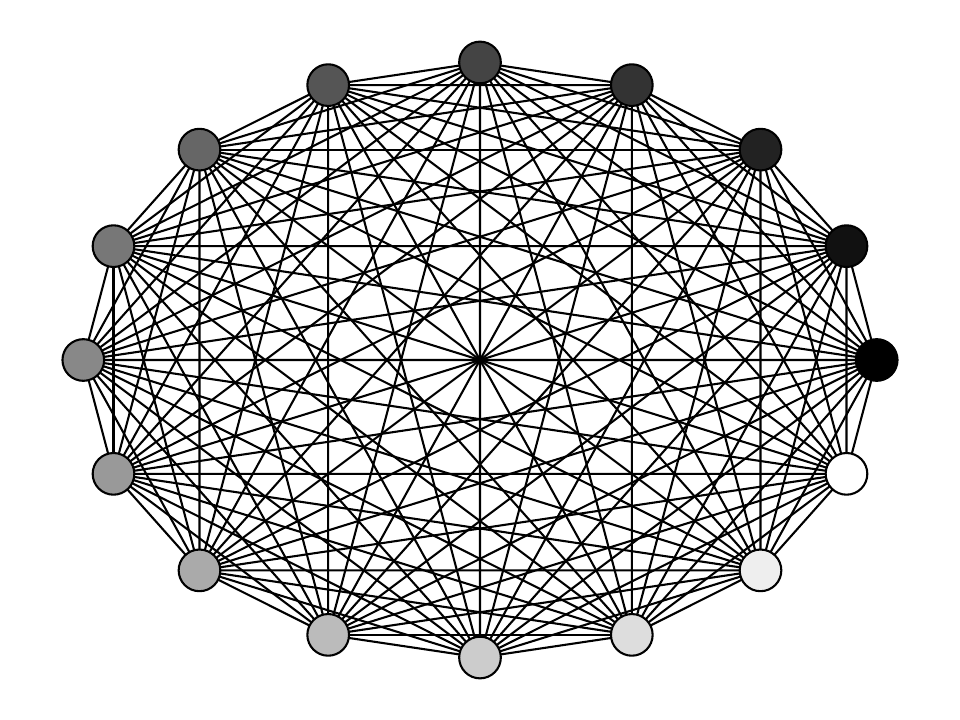}}
    \caption{Comparison between two classical Markov chains. The network in Fig.~\ref{fig:first-neighbor-markov} shows a first-neighbor chain, while the network in Fig.~\ref{fig:full-markov} is fully connected, allowing the walker to move to any of the \(N-1\) remaining nodes.}
    \label{fig:walkers}
\end{figure}

As mentioned, our hybrid-quantum algorithm inspires from the works of Nichol and Austin~\cite{nichol2021improved, austin2021structured} where a classical random walk is implemented to perform the \ac{dm} on discrete categorical data.
Despite our \ac{dtqw} can only make a forward or backward step along the nodes graph, here we decided to compare it against the classical random walk implementations that has been used, which allow the walker to remain within the node $i$ with a certain probability which is directly connected to the scaling factor $\beta.$
In the first formulation also illustrated in the \cref{fig:walkers}, the classical walker moves through the transition matrix that is composed by the adjacency matrix of the ring shape chain with a constant scheduler $\beta = 0.9$ and the identity matrix multiplied by a factor of $1-\beta$, meaning that the walker has a chance of remaining in the same position $i^{\text{th}}$, as following
\begin{equation}
    \mathbf{Q} = (1-\beta)\mathbf{I} + \beta \mathbf{A}
\end{equation} In fact, the effect of the identity matrix is to leave the position of the walker unvaried.

Thus, the corresponding Markov chain in compact form becomes
\begin{equation}
    \pi_t = \Bigl(\prod_{t=1}^T\mathbf{Q}^t\Bigr) \pi_0,
\end{equation}
where $\mathbf{Q}$ is multiplied $t$ times and $\pi_0$ is the initial data distribution, which in this case corresponds to a vector of zeros of length $2^{N_q}$ except for the first element which is set to 1. The corresponding elements take the form
\begin{equation*}
    [Q]_{ij} = \begin{cases}
        1-\beta & \text{if } i=j \\
        \beta/2 & \text{if } i \neq j
    \end{cases}
\end{equation*}
For the second case depicted in \cref{fig:walkers}, which corresponds to the actual set-up implemented in ~\cite{austin2021structured}, the transition matrix is modeled as an identity matrix multiplied by a scheduler that changes according to time, and a fully-connected adjacency matrix. In particular, the walker can diffuse on all $2^{N_q-1}$ the sites, resulting into an accelerated diffusion. The power of this protocol resides in the time scheduler
\begin{equation}
    \beta_t = 1-\frac{\alpha_t}{\alpha_{t-1}},
    \label{scheduler-classic}
\end{equation}
where $\alpha_t = \frac{f(t)}{f(0)}, \ f(t) = \cos(\frac{t/T+s}{1+s}\frac{\pi}{2})^2$ and $s=(\frac{2^{N_q}}{2} - 0.5)^{-1}$ accordingly to~\cite{nichol2021improved}. Finally the time-dependent transition matrix for this set-up is defined as
\begin{equation}
    \mathbf{Q}_t = (1-\beta_t)\mathbf{I} + \beta_t{\mathbb{1}\mathbb{1}^T/K}
\end{equation}
\begin{equation*}
    [Q_t]_{ij} = \begin{cases}
        1-\frac{K-1}{K}\beta_t & \text{if } i=j \\
        \frac{\beta_t}{K} & \text{if } i \neq j.
    \end{cases}
\end{equation*}

\subsection{Discrete-time quantum walk}
\label{sec:quantumWalk}

The unitary operator that allows the walker to perform a single time step is defined as
\begin{equation}
\begin{aligned}
U
&= S \cdot (C \otimes I) = S \cdot (H \otimes I) \\
&=
\Bigl(
    \ket{0}_C\bra{0}_C
    \otimes
    \sum_i \ket{i+1}_P\bra{i}_P
    +
    \ket{1}_C\bra{1}_C
    \otimes
    \sum_i \ket{i-1}_P\bra{i}_P
\Bigr) \\
&\qquad \cdot\left(\frac{1}{\sqrt{2}}\bigl(\ket{0}_C\bra{0}_C+ \ket{0}_C\bra{1}_C+ \ket{1}_C\bra{0}_C- \ket{1}_C\bra{1}_C\bigr)\otimes I\right),
\end{aligned}
\end{equation} where each subscript refers to the corresponding quantum state.
To give a simple example of the details of the \ac{dtqw} algorithm, we consider the global quantum state defined in \cref{globalState}, where the available positions the walker can jump to are only four, $\ket{00}, \ket{01},\ket{10},\ket{11}$ respectively. The details of the dynamic of the quantum walker is described as follows
\begin{equation}
\ket{\Psi} = \ket{s} \otimes \ket{\psi} =
\left(
    s_0 \ket{0}
    + s_1 \ket{1}
\right)
\otimes
\frac{1}{\sqrt{N}}
\Bigl(
    \alpha_{00}\ket{00}
    + \alpha_{01}\ket{01}
    + \alpha_{10}\ket{10}
    + \alpha_{11}\ket{11}
\Bigr),
\end{equation}
where $N = |\alpha_{00}|^2 + |\alpha_{01}|^2 +|\alpha_{10}|^2 +|\alpha_{11}|^2$ is the normalization factor, which is not required for $\ket{s}$ since it is already normalized.
The unitary operation is repeated multiple times to preserve quantum correlations and to allow for different position states to interfere with one another. In this way, we observe a completely different behavior compared to the classical random walk.
Nevertheless, when we measure the coin state we obtain two different cases depending on the state of the coin immediately before the measurement. In particular, we have
\begin{equation}
    U\ket{\Psi} = U(\ket{s}\otimes\ket{\psi}) = S \cdot(H\otimes I)(\ket{s}\otimes\ket\psi).
\end{equation}
Here, we distinguish the application of the shift operator on the entangled state between the coin and the position state, and the action of the Hadamard gate.
The shift operator yields to two distinct cases depending on the coin operator state, in fact
\begin{equation}
\begin{aligned}
&S(\ket{0}\otimes\ket{i}) = \ket{0}\otimes \ket{i+1}, \quad \\
&S(\ket{1}\otimes\ket{i})= \ket{0}\otimes \ket{i-1}.\\
\end{aligned}
\end{equation}
If we combine the effects of ordered sequence of the Hadamard gate already mentioned in \cref{ref:qc} with the shift operator on the position $\ket{00}$, we obtain

\begin{equation}
    \begin{aligned}
        {\displaystyle |{0 }\rangle \otimes \alpha_{00}|00\rangle \;\,{\overset {H}{\longrightarrow }}\;\,{\frac {1}{\sqrt {2}}}(|{0 }\rangle +|{1 }\rangle )\otimes \alpha_{00}|00\rangle \;\,{\overset {S}{\longrightarrow }}\;\,{\frac {\alpha_{00}}{\sqrt {2}}}(|{0 }\rangle \otimes |01\rangle +|{1 }\rangle \otimes |{11}\rangle )},\quad\\
        {\displaystyle |{1 }\rangle \otimes \alpha_{00}\ket{00} \;\,{\overset {H}{\longrightarrow }}\;\,{\frac {1}{\sqrt {2}}}(|{0 }\rangle -|{1 }\rangle )\otimes \alpha_{00}|00\rangle \;\,{\overset {S}{\longrightarrow }}\;\,{\frac {\alpha_{00}}{\sqrt {2}}}(|{01 }\rangle \otimes |1\rangle +|{1 }\rangle \otimes |{11}\rangle ),}
    \end{aligned}
\end{equation}
where we have exploited the periodic boundary conditions ($\ket{00}\xrightarrow{\ket{i} \rightarrow{\ket{i+1}}}\ket{01}$, and $\ket{00}\xrightarrow{\ket{i} \rightarrow{\ket{i-1}}}\ket{11}$).

\subsection{Additional results}
\subsubsection{BloodMnist}

We display 100 original and reproduced BloodMNIST images with \ac{dtqw} and the classical Markov chain at $\beta$ fixed and time dependent respectively. 
\begin{figure}[htbp]
    \centering
    \subfloat[Original.]{%
        \includegraphics[width=0.5\linewidth]{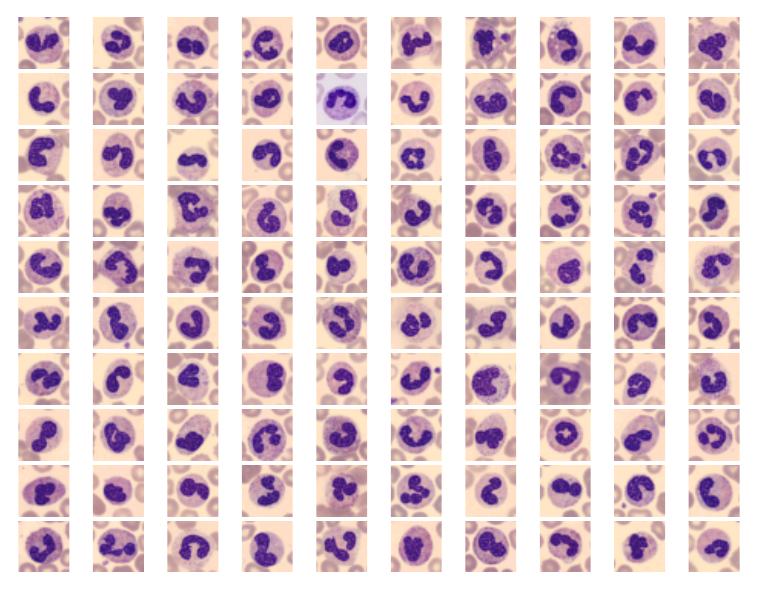}}
    \hfill 
    \subfloat[Generated with the \ac{dtqw} executed on IBM quantum device.]{%
        \includegraphics[width=0.5\linewidth]{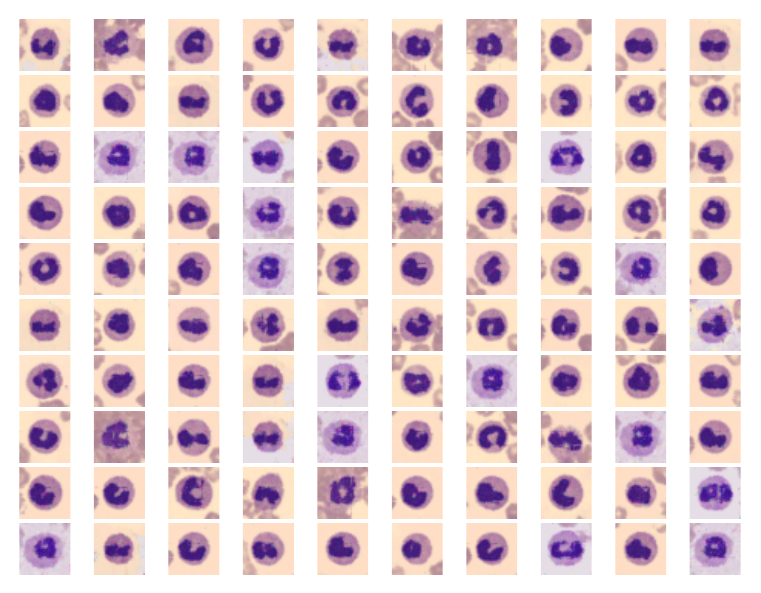}}
        \hfill
    \subfloat[Generated with the classical Markov chain with fixed scheduler $\beta$.]{%
        \includegraphics[width=0.5\linewidth]{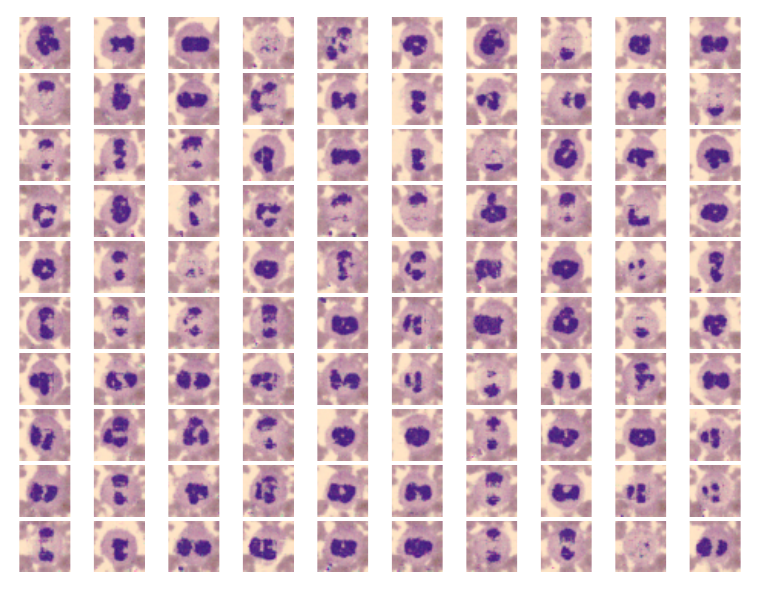}}
    \hfill
    \subfloat[Generated with the classical Markov chain with the variable time-dependent scheduler $\beta_t$.]{%
        \includegraphics[width=0.5\linewidth]{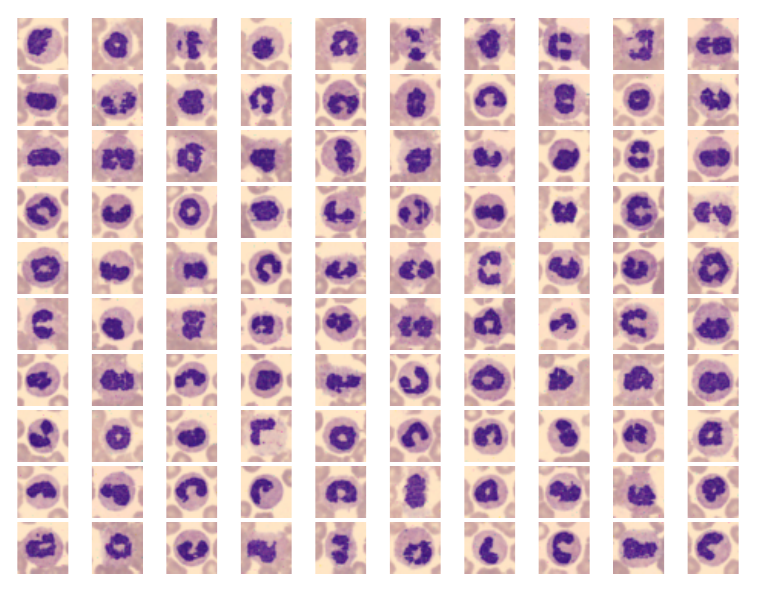}}
    \caption{Batch of 100 random BloodMNIST images.}
    \label{fig:examplesBlood}
\end{figure}

The following figures depict $6$ consecutive time-steps of the model dynamics, in particular $T = \{0, \  1, \ 2, \ 10, \ 20, \ 30\}$ for the BloodMNIST and BraTS2020 images. We decide to not display the same dynamics for the volume data since after a single diffusion time-step it would not be possible to visualize the volume rendering of the region that gets diffused in time. Respectively, the top row shows a batch of 9 samples that is progressively diffused until $T=30$ where converges to the uniform distribution. Below, we draw the probability distribution function associated to the batch of samples, that is decomposed into its three independent channel components. The third row displays the backward dynamics, in which the model learns to denoise the images and to reversely generate novel samples.
Following the same logic, the last row shows the denoising trend of the three distinct channels of the probability distribution function.

\begin{figure}[th!]
\includegraphics[width=\textwidth]{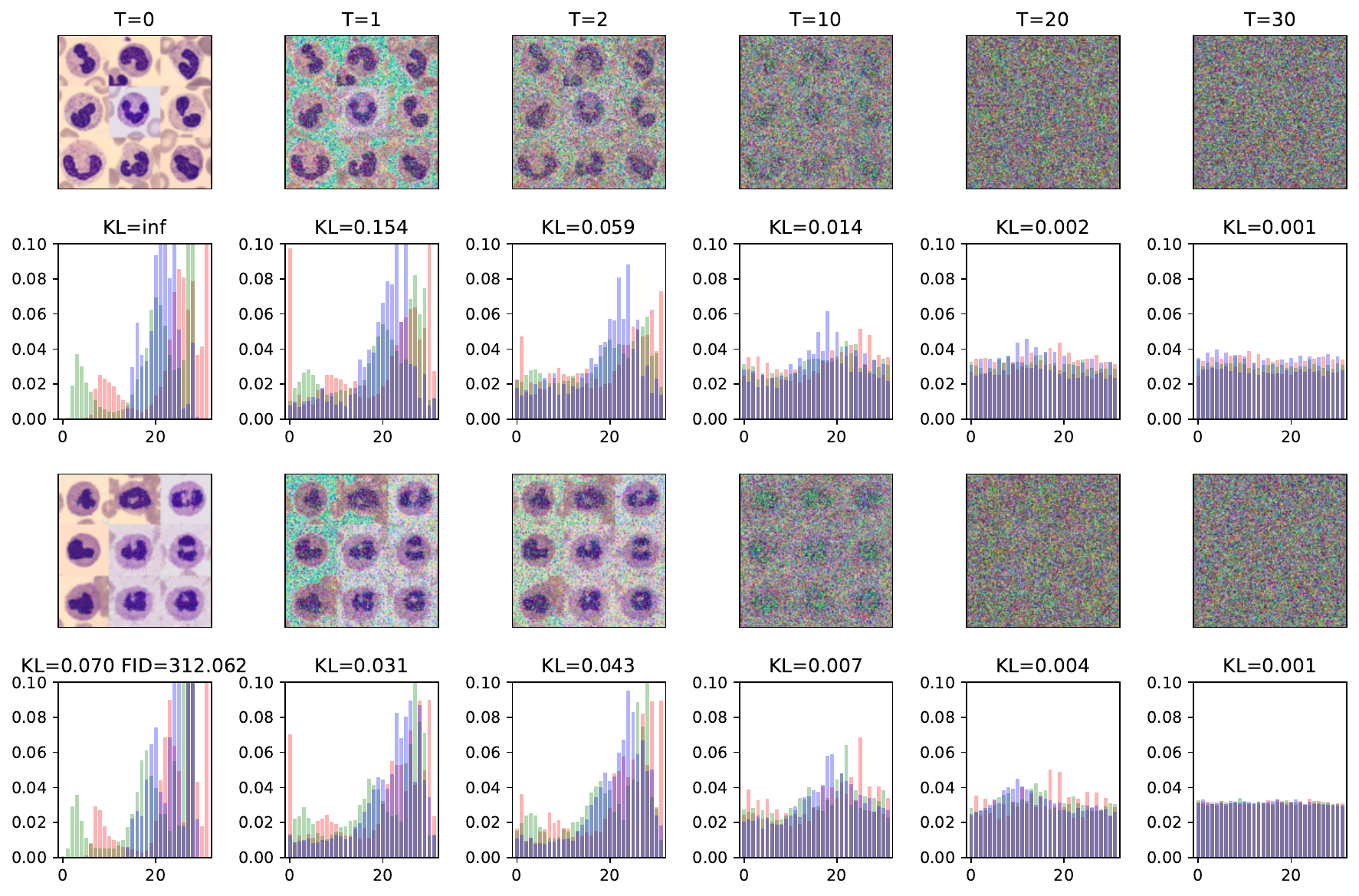}
    \caption{Diffusion process executed on IBM Torino for $T=30$ steps using the \ac{dtqw} on BloodMNIST with 32 states (top) and
    backward process performed by the U-Net trained for $\sim 20\,000$ epochs (bottom). We show the evolution for 9 random combined samples of original and reconstructed images and the evolution of the pixel-value distribution for the three channels (in red, green and blue). The visualized target distribution is the Uniform for the forward and the one of the original dataset for the backward. We also report the \ac{kl} divergence between the data and target distributions and the final \ac{fid} of the final generated images.}
    \label{fig:imageBlood}
\end{figure}

\begin{figure}[ht!]
    \centering
    \includegraphics[width=\linewidth]{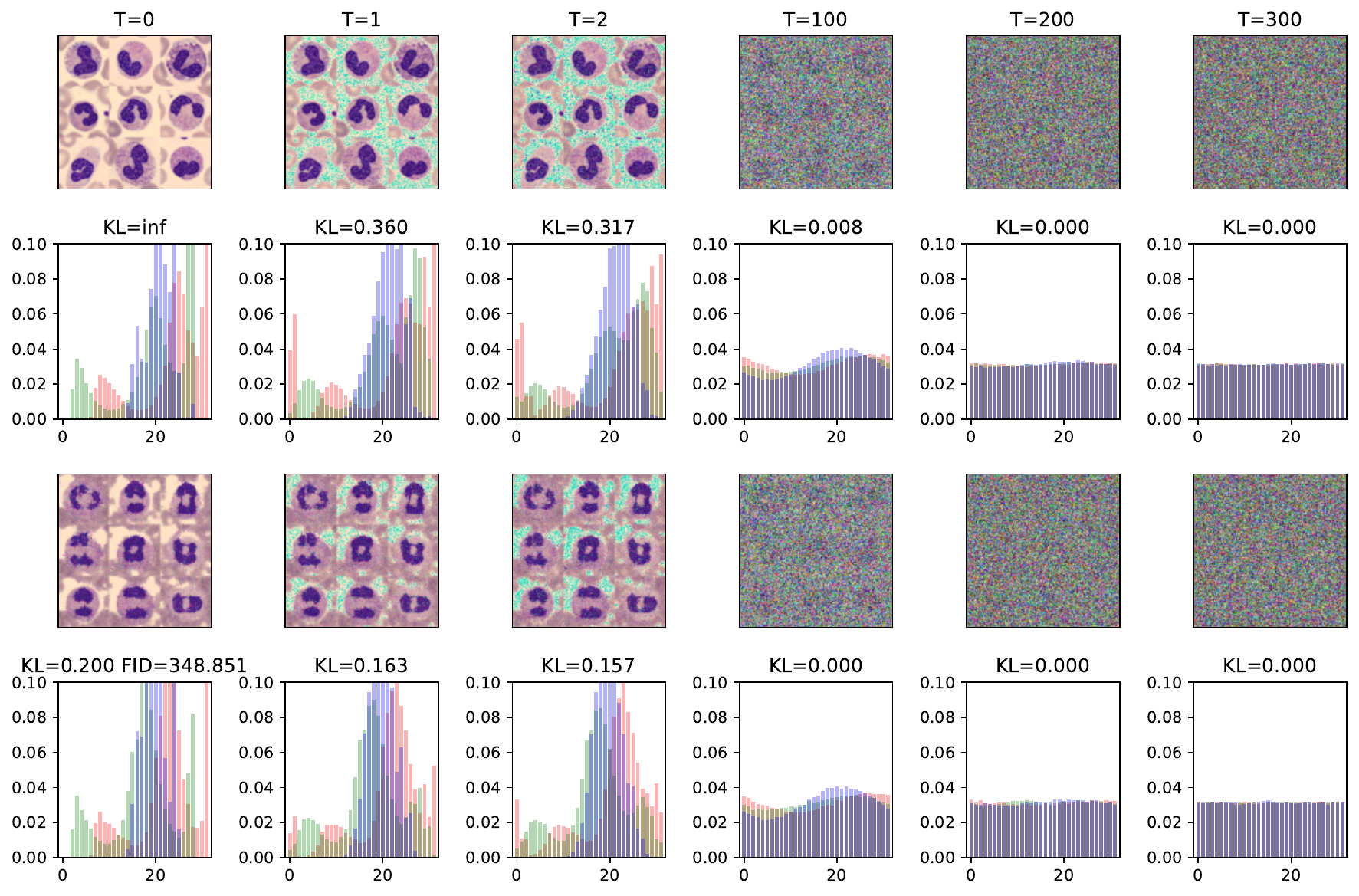}
    \caption{Diffusion process performed with the classical Markov chain on a 32-node ring network embedding pixel intensity levels with nearest-neighbor connection of RGB BloodMNIST images (Top) and backward process performed by the U-Net trained for $\sim 20 \ 000$ epochs (Down). The $\beta$ scheduler is kept fixed $\beta=0.9.$}
    \label{fig:betaFixed-cdm}
\end{figure}

\begin{figure}[ht!]
    \centering
    \includegraphics[width=\linewidth]{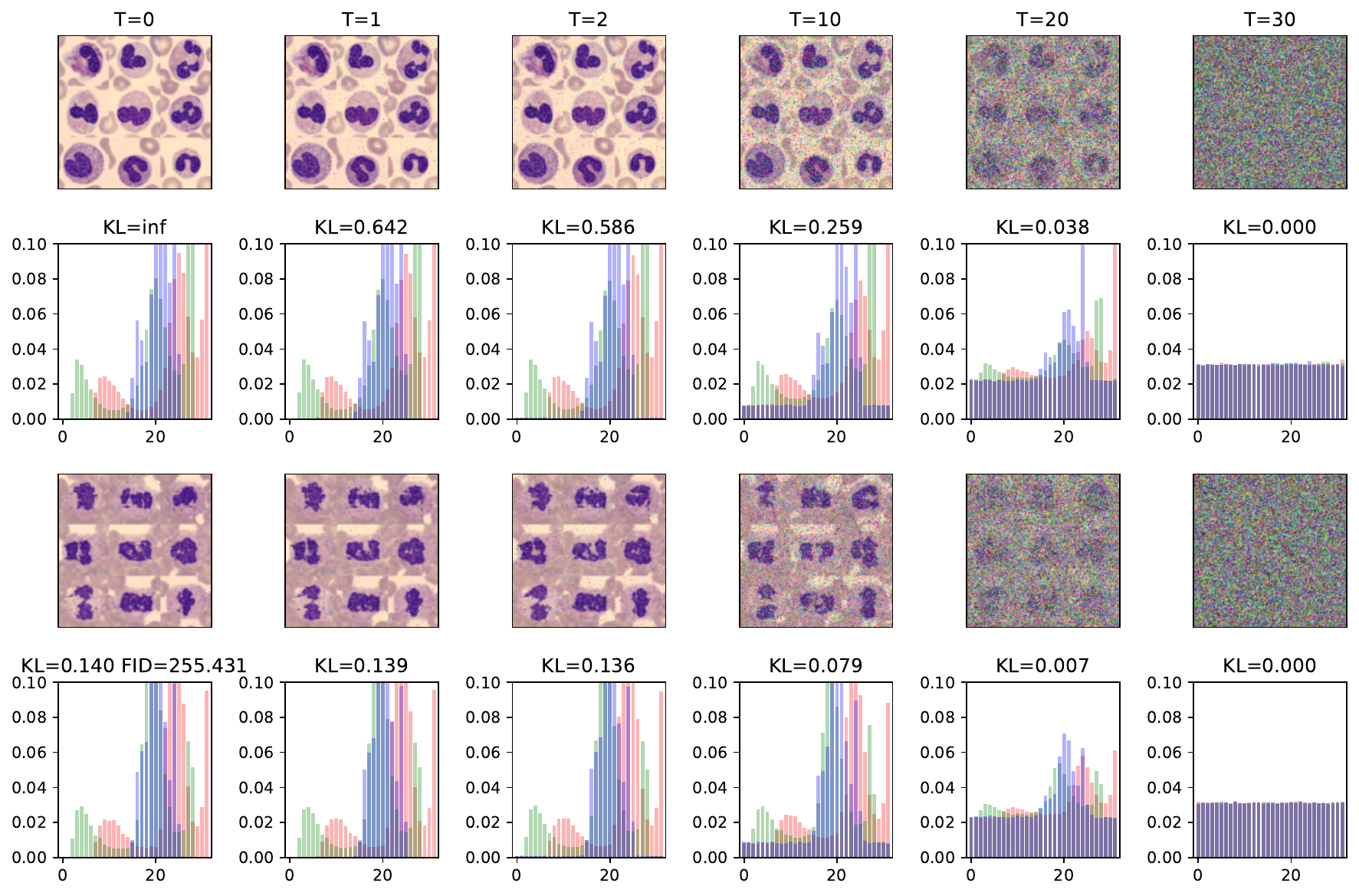}
    \caption{Diffusion process performed with the classical Markov chain on a 32-node ring network embedding pixel intensity levels with full connection of RGB BloodMNIST images (Top) and backward process performed by the U-Net trained for $\sim 20 \ 000$ epochs (Down). The $\beta$ scheduler varies at each time step according to  Eq. \ref{scheduler-classic}.}
    \label{fig:betaVariable-cdm}
\end{figure}

We show the results for the classical Markov chain for two distinct set-ups introduced in \ref{classicalDiffusion}.
In the first, we diffuse and train the backward model by using a transition matrix composed of the probabilities obtained by using a fixed $\beta$ scheduler function and a closed-chain graph topology with first-neighbor connections. In the third picture we display the same dynamics for the classical Markov chain with time-dependent $\beta$ scheduler function and a fully connected graph topology. We repeat the same scheme for the BraTS2020 dataset in the subsequent subsection.

\vspace{2cm}
\newpage
\newpage
\subsubsection{BraTS2020}

We display 100 original and reproduced BraTS2020 images with \ac{dtqw} and the classical Markov chain at $\beta$ fixed and time dependent respectively.

\begin{figure}[htbp]
    \centering
    \subfloat[Original.]{%
        \includegraphics[width=0.5\linewidth]{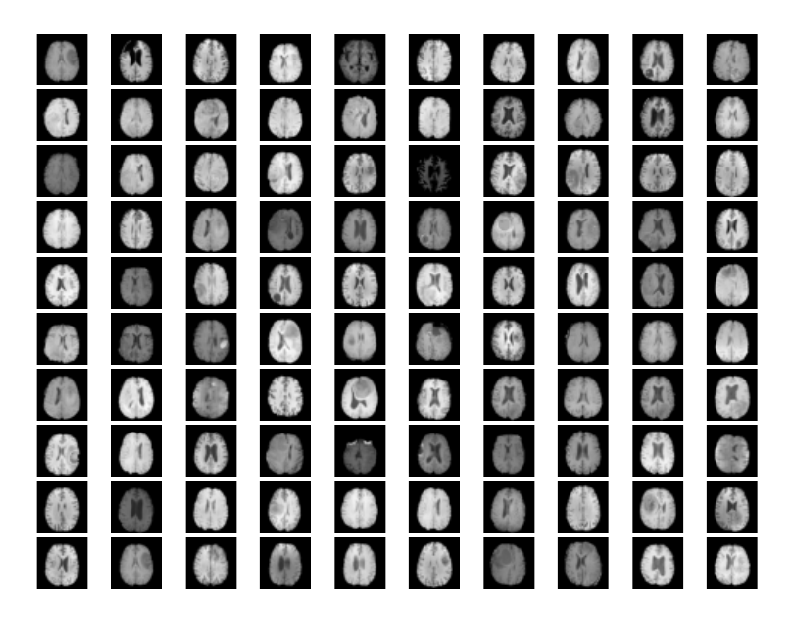}}
    \hfill 
    \subfloat[Generated with the \ac{dtqw} executed on IBM quantum device.]{%
        \includegraphics[width=0.5\linewidth]{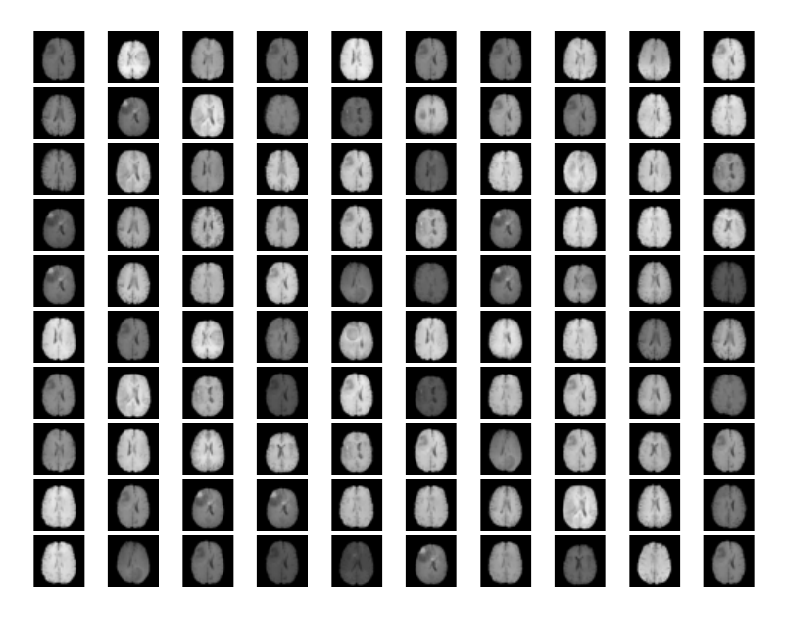}}
        \hfill
    \subfloat[Generated with the classical Markov chain with fixed scheduler $\beta$.]{%
        \includegraphics[width=0.5\linewidth]{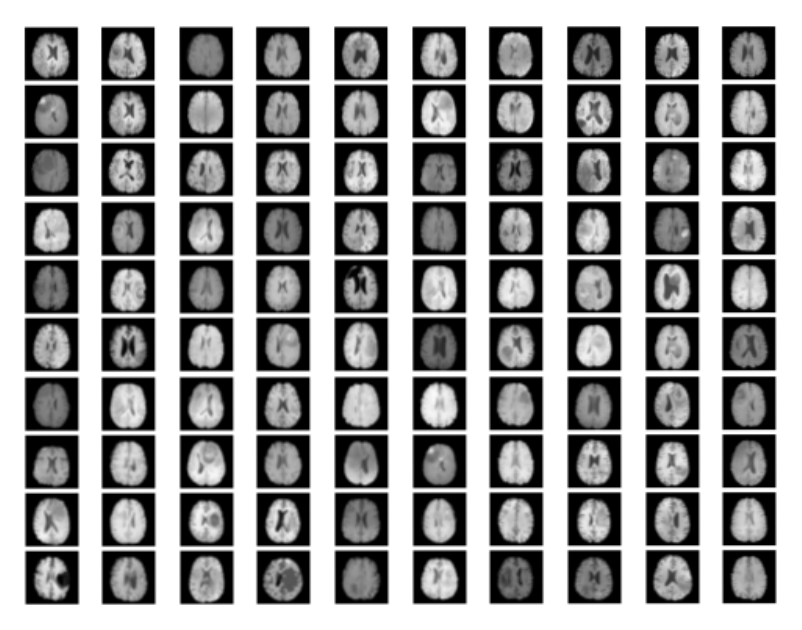}}
    \hfill
    \subfloat[Generated with the classical Markov chain with the variable time-dependent scheduler $\beta_t$.]{%
        \includegraphics[width=0.50\linewidth]{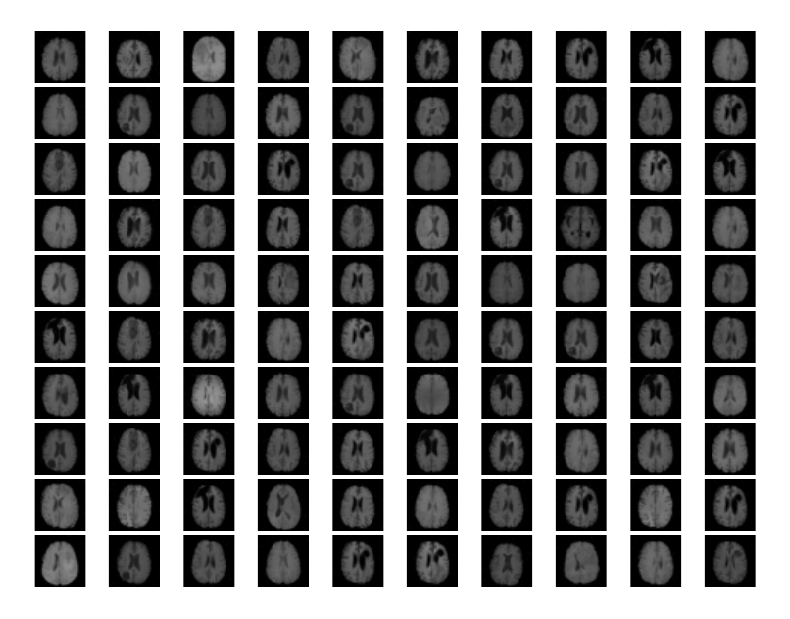}}
    \caption{Batch of 100 random BraTS2020 images.}
    \label{fig:examplesBrain}
\end{figure}

\newpage

\begin{figure}[th!]
    \centering
    \includegraphics[width=\textwidth]{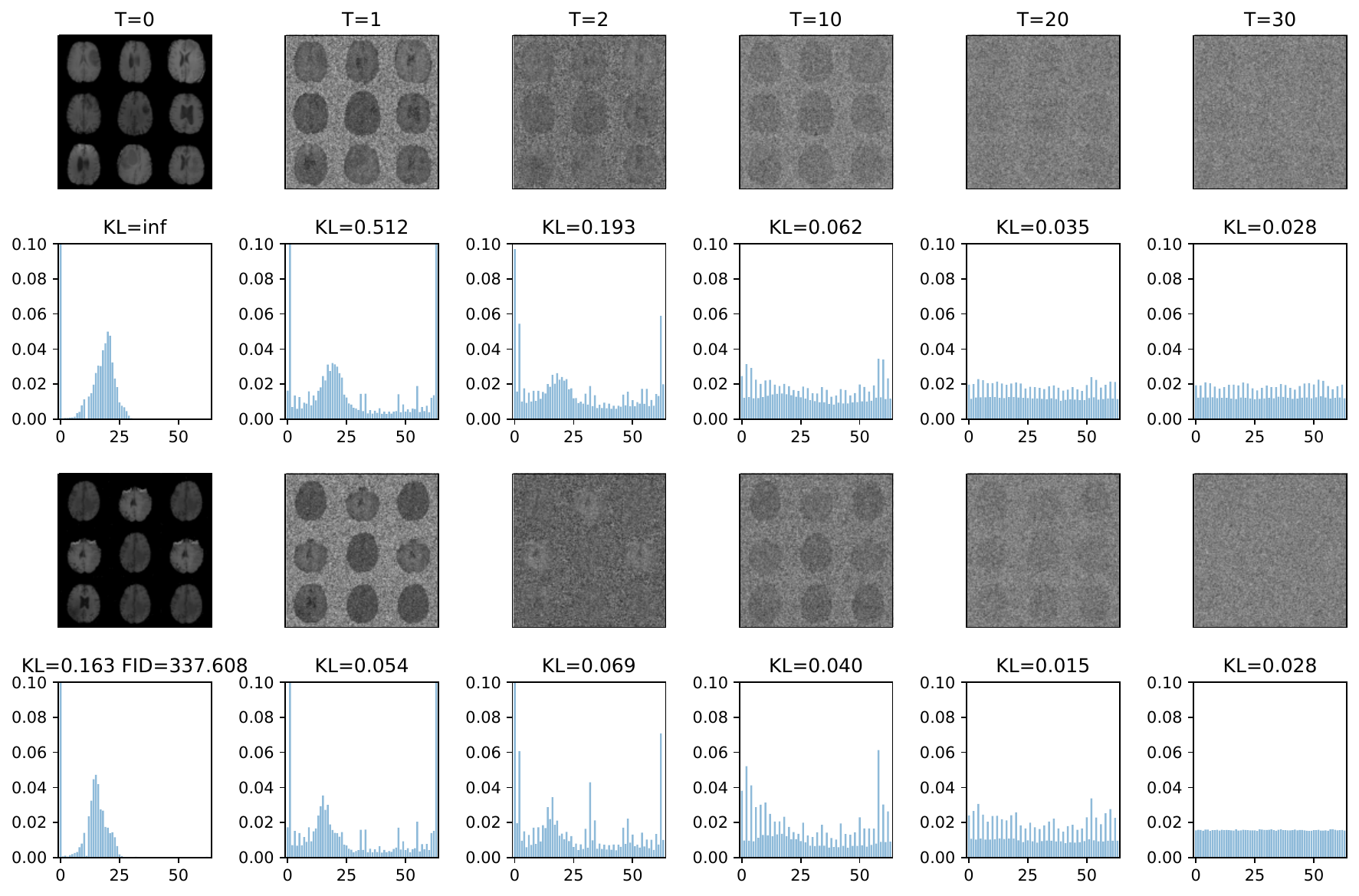}
        \caption{Diffusion process executed on IBM Torino for $T=30$ steps using the \ac{dtqw} on grayscale BraTS2020 brain images of 64 states. The top row shows the forward process, while the bottom row shows the backward process performed by the U-Net trained for $\sim 20\,000$ epochs. We show the evolution of 9 random combined samples of original and reconstructed images and the evolution of the pixel-value distribution compared with the target distribution.}
    \label{fig:imageBrain-dtqw}
\end{figure}

\begin{figure}[th!]
\centering
    \includegraphics[width=\textwidth]{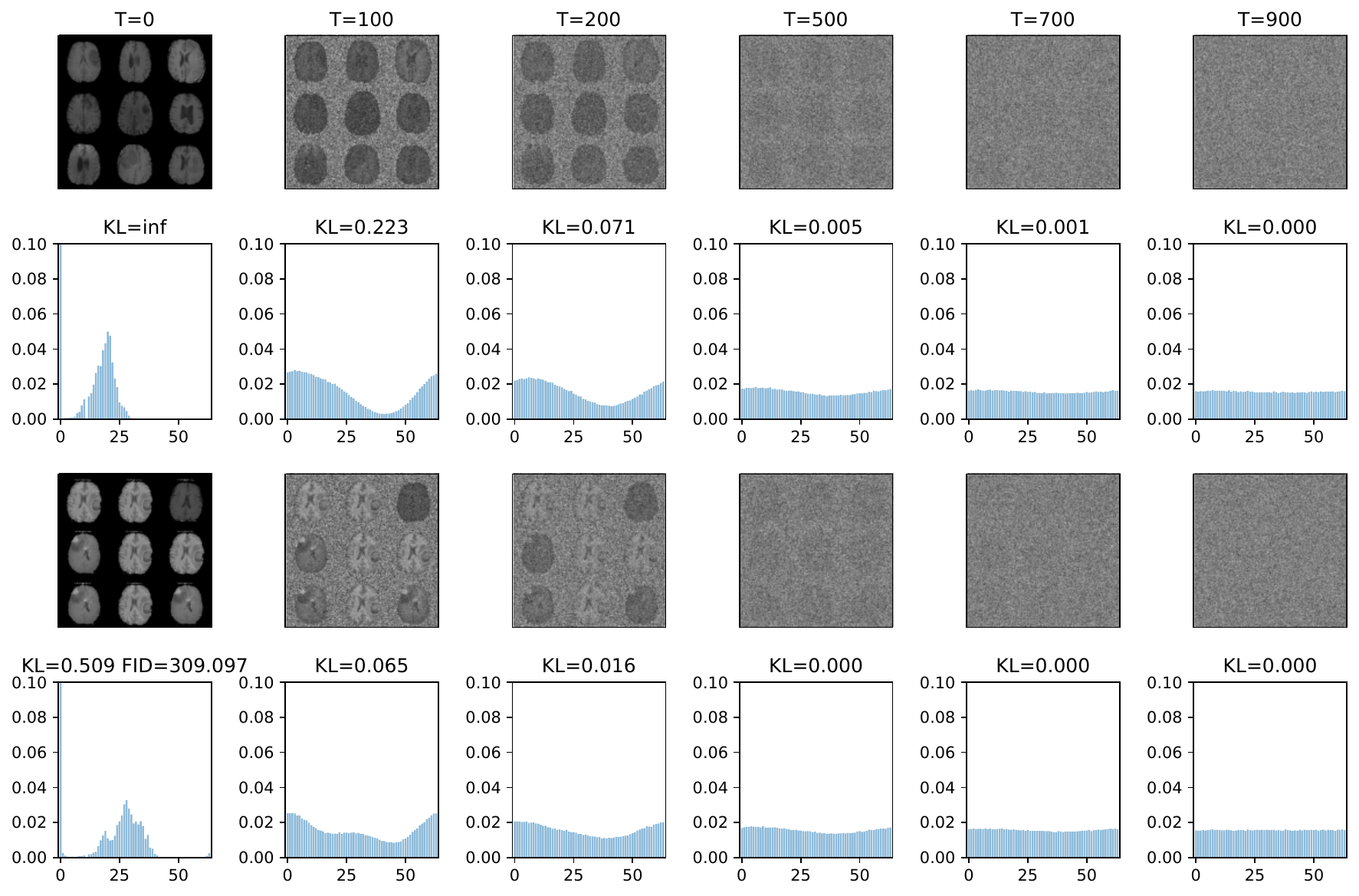}
        \caption{Diffusion process performed with the classical Markov chain on a 64-node ring network embedding pixel intensity levels with nearest-neighbour connections. The top row shows the forward process, while the bottom row shows the backward process performed by the U-Net trained for $\sim 20\,000$ epochs. The $\beta$ scheduler is set to 0.9 with $T=900$ steps, reaching a uniform distribution.}
    \label{fig:imageBrain-classical-sub}
\end{figure}

\begin{figure}[th!]
\includegraphics[width=\textwidth]{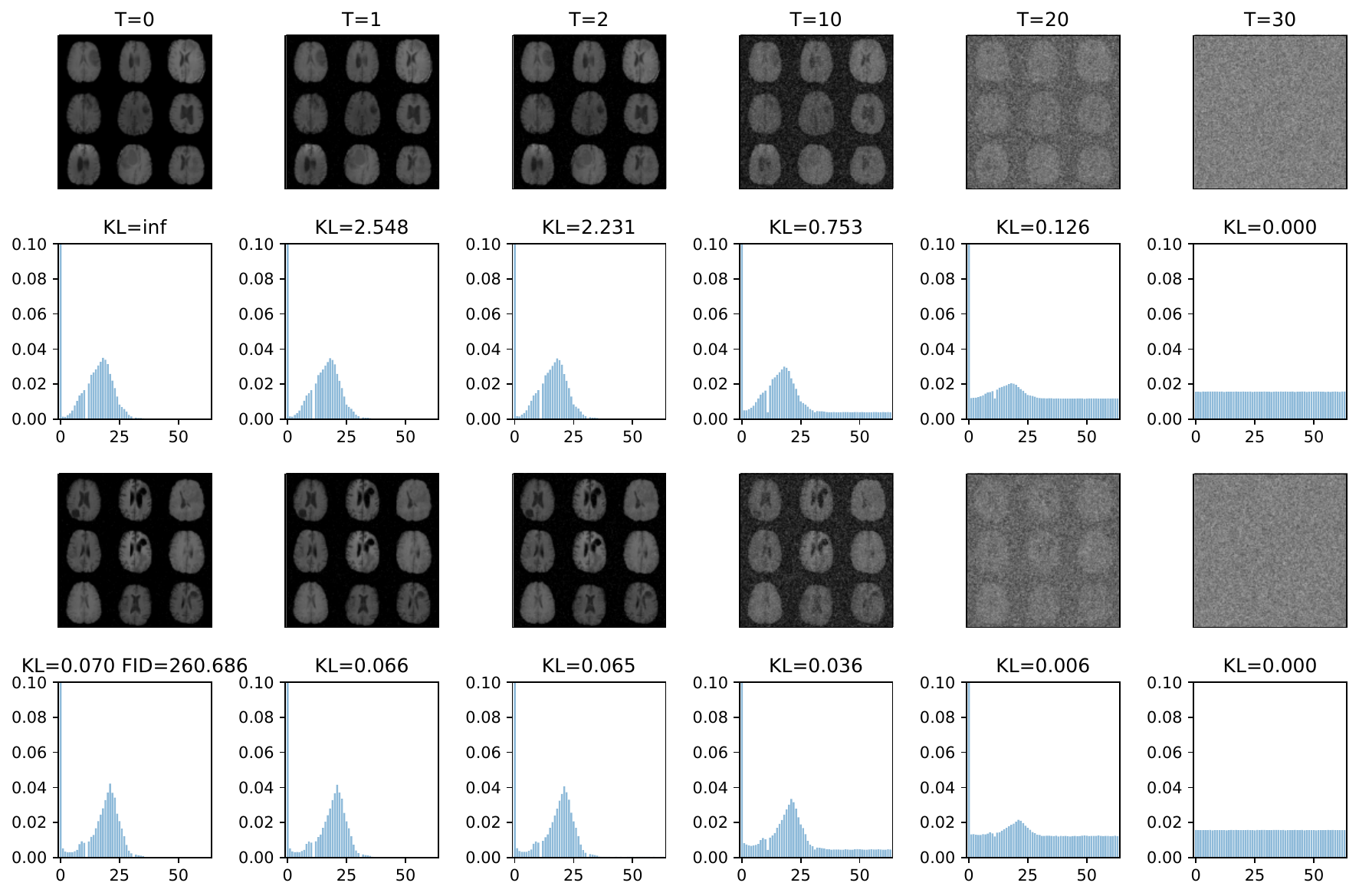}
    \caption{Diffusion process performed with the classical Markov chain on a 64-nodes ring fully-connected network embedding pixel intensity levels (Top) and backward process performed by the U-Net trained for $\sim 20\,000$ epochs (Down). The $\beta$ scheduler is $\beta_t$ and requires only 30 time steps to converge.}
    \label{fig:imageBrain-classical-betaVariable}
\end{figure}
\newpage
\newpage

\subsubsection{PCA on both models}
Eventually, we display the two largest eigenvalues obtained by the PCA applied to the generated data, by order BloodMNIST, BraTS2020 and FractureMNIST respectively, for both of the distinct forward dynamics we executed: quantum on IBM Torino, classical nearest neighbor connection Markov chain with $\beta=0.99$, and the classical Markov chain on a fully connected graph with time-dependent $\beta$.

\begin{figure}[ht]
    \centering
    \subfloat[(Maroon) PCA projection of the classical DM with nearest-neighbor topology and $\beta=0.99$ (100). (Purple) PCA projection of the original data samples (3000).
    \label{fig:pca_blood_betaFixed}]{%
    \includegraphics[width=0.5\textwidth]{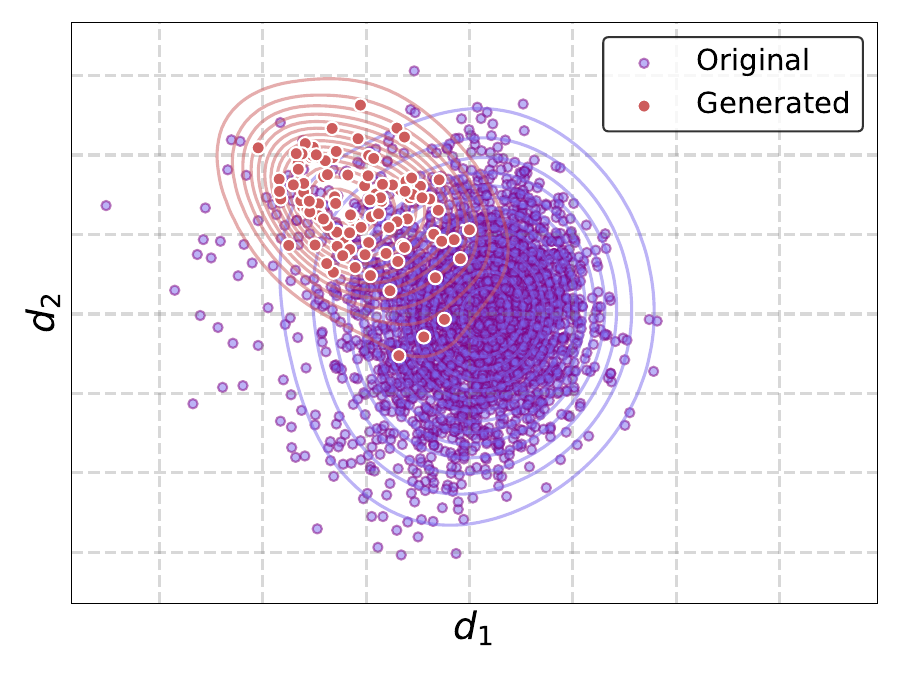}}
    \hfill
    \subfloat[(Orange) PCA projection of the classical DM with fully connected topology and $\beta=\beta_t$ (100). (Purple) PCA projection of the original data samples (3000).
    \label{fig:pca_blood_betaVar}]{%
    \includegraphics[width=0.5\textwidth]{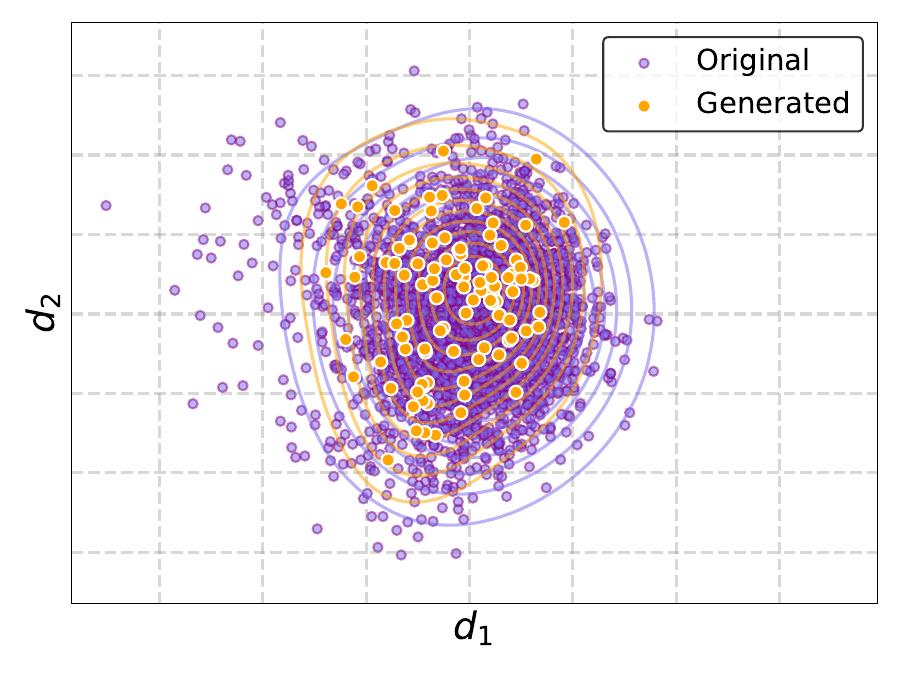}}
    \caption{PCA projections on the BloodMNIST dataset comparing the two distinct versions of the classical DMs. }
\label{fig:pcaBlood_comparison_allmodels}
\end{figure}

\begin{figure}[ht!]
\centering
\subfloat[(Maroon) PCA projection of the classical DM with nearest-neighbor topology and $\beta=0.99$ (100). (Purple) PCA projection of the original data samples (450).
    \label{fig:pca_brain_betaFixed}]{%
    \includegraphics[width=0.5\textwidth]{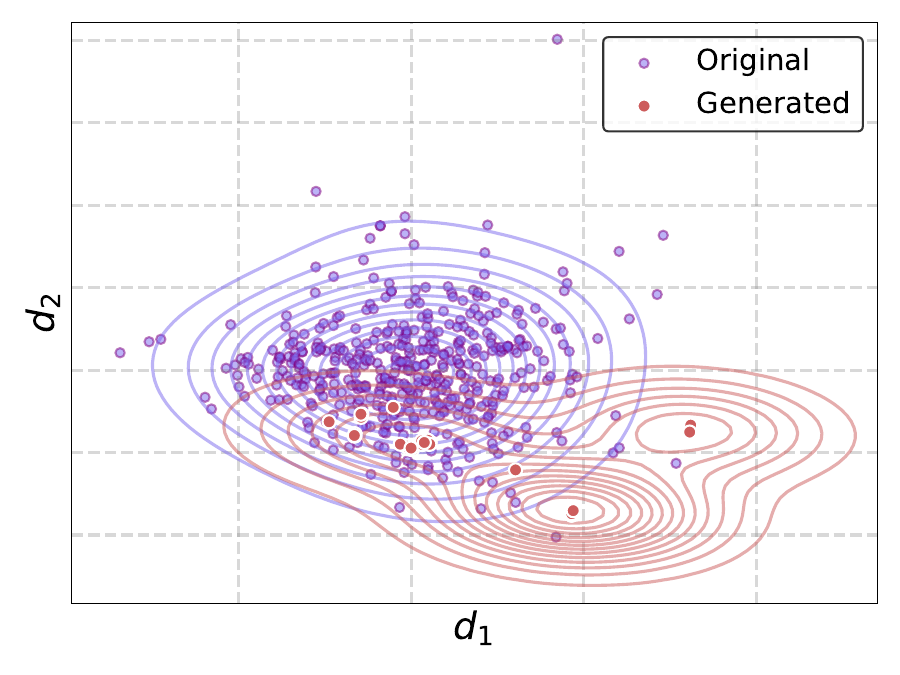}}
\hfill
\subfloat[(Orange) PCA projection of the classical DM with fully connected topology and $\beta=\beta_t$ (100). (Purple) PCA projection of the original data samples (450).
    \label{fig:pca_brain_betaVar}]{%
    \includegraphics[width=0.5\textwidth]{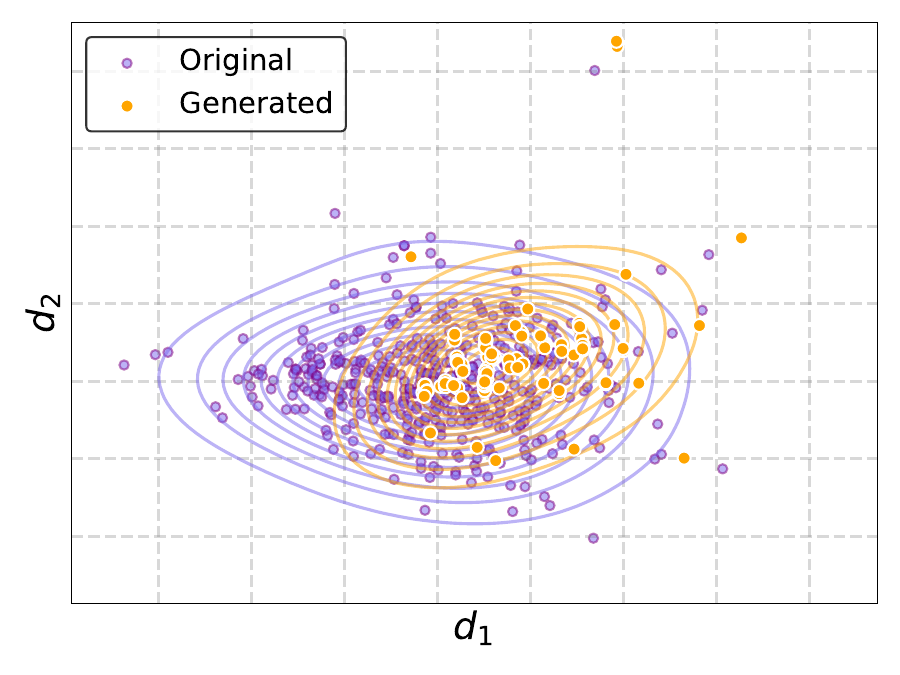}}
\caption{PCA projections on the BraTS2020 dataset comparing the two distinct versions of the classical DMs.}
\label{fig:pcaBrain_comparison_allmodels}
\end{figure}

\begin{figure}[ht!]
\centering

\subfloat[(Maroon) PCA projection of the classical DM with nearest-neighbor topology and $\beta=0.99$ (100). (Purple) PCA projection of the original data samples (1370).
    \label{fig:pca_fract_betaFixed}]{%
    \includegraphics[width=0.5\textwidth]{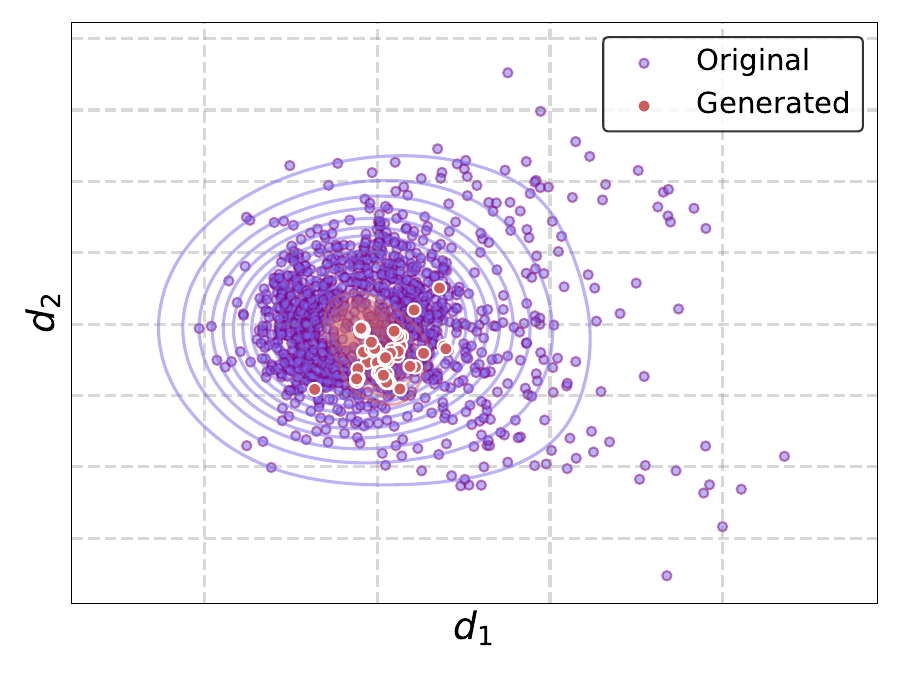}}
\hfill
\subfloat[(Orange) PCA projection of the classical DM with fully connected topology and $\beta=\beta_t$ (100). (Purple) PCA projection of the original data samples (1370).
    \label{fig:pca_fract_betaVar}]{%
    \centering
    \includegraphics[width=0.5\textwidth]{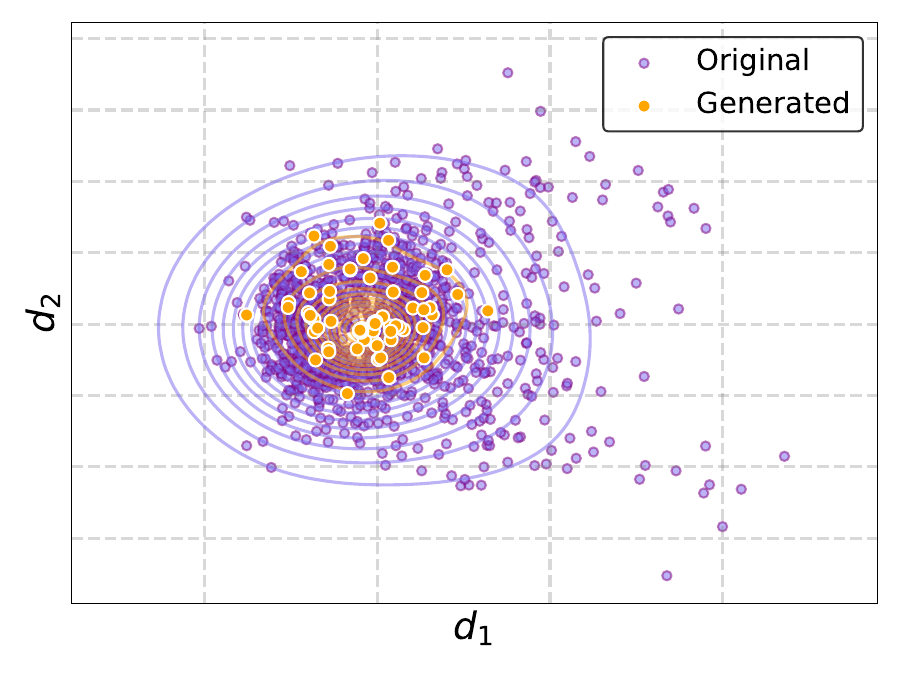}}
\caption{PCA projections on the FractureMNIST dataset comparing the two distinct versions of the classical DMs.}
\label{fig:pcaFract_comparison_allmodels}

\end{figure}

\end{document}